\documentclass[aps, pra, reprint, amsmath,amssymb,graphicx, longbibliography]{revtex4-1}

\usepackage{hyperref}

\usepackage{subfigure}
\usepackage{mathtools}
\usepackage{amssymb}
\usepackage{amsfonts}
\usepackage{amsmath}
\usepackage{amsthm}
\usepackage{latexsym}
\usepackage{dcolumn}

\usepackage{printlen}

\usepackage{amsmath}
\usepackage{braket}
\usepackage{graphicx}

\newcommand{\appropto}{\mathrel{\vcenter{
  \offinterlineskip\halign{\hfil$##$\cr
    \propto\cr\noalign{\kern2pt}\sim\cr\noalign{\kern-2pt}}}}}

\DeclareMathOperator*{\argmax}{arg\,max}

\begin{document}
\title{Encoding a Qubit into a Cavity Mode in Circuit-QED using Phase Estimation}

\author{B.M. Terhal and D. Weigand}
\affiliation{JARA Institute for Quantum Information, RWTH Aachen University, 52056 Aachen, Germany}
\date{\today}

\begin{abstract}
%DW RESUBMIT CHANGE: The new simulation gives 94\% for 8 rounds of ape
Gottesman, Kitaev and Preskill have formulated a way of encoding a qubit into an oscillator such that the qubit is protected against small shifts (translations) in phase space. The idea underlying this encoding is that error processes of low rate can be expanded into small shift errors. The qubit space is defined as an eigenspace of two mutually commuting displacement operators $S_p$ and $S_q$ which act as large shifts/translations in phase space. We propose and analyze the approximate creation of these qubit states by coupling the oscillator to a sequence of ancilla qubits. This preparation of the states uses the idea of phase estimation where the phase of the displacement operator, say $S_p$, is approximately determined. We consider several possible forms of phase estimation. We analyze the performance of repeated and adapative phase estimation as the simplest and experimentally most viable schemes given a realistic upper-limit on the number of photons in the oscillator. We propose a detailed physical implementation of this protocol using the dispersive coupling between a transmon ancilla qubit and a cavity mode in circuit-QED. We provide an estimate that in a current experimental set-up one can prepare a good code state from a squeezed vacuum state using $8$ rounds of adapative phase estimation, lasting in total about $4 \mu$ sec., with $94\%$ (heralded) chance of success.   
\end{abstract}
\maketitle

%\tableofcontents

\section{Introduction}
\label{sec:intro}
It is to be expected that quantum error correction will be essential in the implementation of reliable, large-scale, quantum computation. Efforts are underway for superconducting transmon qubits \cite{koch+:transmon} coupled to microwave-resonators (circuit-QED) to realize components of a surface code architecture in which robust logical qubits are comprised of many $O(10-100)$ elementary transmon qubits and resonators \cite{DS:science, fowler:practical}.
As the qubit and resonator overhead of such encoding is rather daunting and inefficient, one may ask whether there exist more efficient ways of using high-coherence qubits and long life-time microwave resonators (see some numbers in Table \ref{table:parameters}) to encode robust logical qubits. In this paper we will explore the proposal by Gottesman, Kitaev and Preskill \cite{GKP} to encode a qubit into an oscillator and take a first stab at analyzing how a qubit can be encoded and preserved in a single mode of a microwave cavity. This scheme may not only be interesting as an alternative route towards scalable quantum computation, but also as a way of generating a highly non-classical state in a (microwave) cavity which one can actively protect against photon loss. \\

The GKP (Gottesman-Kitaev-Preskill) code is of interest as it offers the possibility to use the cavity mode to store a single qubit, while a transmon qubit coupled to this cavity mode can be used to manipulate the state of the cavity.
The goal is to prepare and preserve a qubit in a cavity mode such that it has a coherence time much longer than that of the cavity or the transmon qubit itself. The GKP scheme is attractive in that gates on this qubit are relatively straightforward as Clifford gates can be realized using linear optical componenents, see \cite{GKP} and Appendix \ref{sec:phase}. In addition, this qubit-into-oscillator scheme can be concatenated with a surface code or an other stabilizer code for scalable protection (see e.g. \cite{BMT:review} for a surface code with  oscillators).

However, a realistic scheme for preparing the encoded states and performing quantum error correction has been missing so far (in Section \ref{sec:prev} we list some of the previous work) and this is the focus of the present paper. In the remainder of this section we review and discuss the code states and their highly-nonclassical properties. We argue why, at a heuristic level, using phase estimation can provide a means for approximately generating such quantum states. In Section \ref{sec:shift} we review and introduce some tools and formalism which are useful in assessing the quality of a preparation protocol, using a model of displacement or shift errors. In Section \ref{sec:phys_noise} we physically motivate the shift error model by discussing how physical errors can be expanded into such shifts. The shift error model plays a similar role as a Pauli error model for qubits. This section partially reviews some of the insights of \cite{GKP}; it adds to these by showing in Section \ref{sec:phys_shift} how an expansion into shift errors can lead to a reduction in error rate by several orders of magnitude.
Section \ref{sec:prep} is devoted to exploring and understanding how to prepare an approximate code state using the protocol of phase estimation, in this case determining the eigenvalue of a unitary displacement operator. There are many ways to do phase estimation leading to different approximations  for the code states and using different numbers of photons. These phase estimation protocols all consist of a coupling a sequence of single qubit ancillas to an oscillator mode sequentially and measuring the ancilla qubit. We argue why we focus on two simplest protocols: one is repeated phase estimation without feedback (Section \ref{RPE}) and a second one is a simple form of adaptive phase estimation with feedback (Section \ref{sec:feedback}). For those two schemes we explicitly show in Section \ref{sec:num} how well with a small number of rounds/ancilla qubits an approximate code state can be prepared.
We also consider the average number of photons used and the variance thereof, as these numbers are relevant in any experimental realization of the protocol. 

In Section \ref{sec:prop} we discuss a realization of the code state preparation protocol in an experimental circuit-QED set-up where the oscillator corresponds to a cavity mode. We review the circuit-QED transmon-cavity coupling and the physical strength of various parameters as reported in recent experiments in circuit-QED.
We discuss several aspects of an implementation such as the use of a tunable coupling and the implementation of a (transmon) qubit controlled-displacement gate by means of a controlled-rotation or a direct pulse-driven controlled-displacement (details in Appendix D). In a final Section \ref{sec:noise} we discuss at a qualitative level the dominant sources of inaccuracy in implementing the protocol; we especially note that the presence of cavity nonlinearities may require a deeper analysis. We end the paper with a discussion and outlook for future work.

In Appendix A and B we collect some background technical results which are relevant for implementing gates on the code states and performing quantum error correction. Appendix C gives background on the choice of feedback phases in the phase estimation protocol in Section \ref{sec:feedback}.

\subsection{GKP Code States}

In order to introduce the code states, we assume a harmonic oscillator with annihilation operator (creation operator) $a$ ($a^{\dagger}$) with which one can define (dimensionless) quadrature operators $p=\frac{i}{\sqrt{2}}(a^{\dagger}-a)$ (momentum) and $q=\frac{1}{\sqrt{2}}(a+a^{\dagger})$ (position) obeying the canonical commutation relations $[q,p]=i$.
The ideal GKP code is simply a prescription of a two-dimensional subspace in the infinite-dimensional oscillator space. This subspace, called the code space,  is defined as the $+1$ eigenspace of two commuting displacement operators $S_p=e^{-i 2\sqrt{\pi}p}$ and $S_q=e^{i 2\sqrt{\pi}q}$ (One can verify the commutation of these operators by using the identity $e^A a^B=e^B e^A e^{[A,B]}$ for $A,B$ linear combinations of $p$ and $q$). The space fixed by $S_p$ and $S_q$ is two-dimensional as there are additional operators which we can identify as $X$ and $Z$ which commute with both $S_p$ and $S_q$, but which mutually anti-commute.

States in this two-dimensional subspace will be called code states as they encode a qubit. Of course, there are many possible ways of choosing a two-dimensional subspace of a harmonic oscillator to define a qubit, e.g. choose two energy eigenlevels or pick two orthogonal cat states to define $\ket{0}$ and $\ket{1}$. The subspace of the GKP code is one in which $\ket{0}$ and $\ket{1}$ are highly non-classical states and is chosen such that small phase space displacements can be undone, by quantum error correction. In this sense the code offers the possibility to realize a long-lived, well-protected qubit.

What is immediately interesting about a state in this subspace is that the condition $S_p=1$ and $S_q=1$ fixes {\em both} momentum $p=0 \mod \sqrt{\pi}$ and position $q=0\mod \sqrt{\pi}$ to be sharply determined. A code state thus escapes Heisenberg's uncertainty principle by being not localized at a single $ p$ and $q$ but being a superposition of many equally spaced sharp values for $p$ and $q$. Due to having low variance in {\em both} quadratures, we can expect that a code state is useful in metrology for detecting small displacements which shift $p$ or $q$ by less than $\sqrt{\pi}$. 

If we write the operators $S_p$ and $S_q$ as displacements, one has $S_p=D(\sqrt{2\pi})$ and $S_q=D(i\sqrt{2\pi})$ (using $D(\alpha)=\exp(\alpha a^{\dagger}-\alpha^* a)$ with $D(\alpha) \ket{\rm vac}=\ket{\alpha}$ and $\ket{\alpha}$ a coherent state). In this code space one can define the qubit states $\ket{0}$ and $\ket{1}$ as
\begin{align} 
\ket{0} &\propto \sum_{t=-\infty}^{\infty} D(t \sqrt{2\pi}) \ket{q=0}=\sum_{t=-\infty}^{\infty}S_p^t \ket{q=0},  \notag \\
\ket{1} &\propto \sum_{t=-\infty}^{\infty} D(t \sqrt{2\pi}) \ket{q=\sqrt{\pi}}=\sum_{t=-\infty}^{\infty} S_p^t \ket{q=\sqrt{\pi}}.
\label{eq:perfect}
\end{align}
Thus, $\ket{0}$ is a uniform superposition of eigenstates of position $q$ at {\em even} integer multiples of $\sqrt{\pi}$, while $\ket{1}$ is a uniform superposition of eigenstates of position $q$ at {\em odd} integer multiples of $\sqrt{\pi}$.  One can also consider these states as superpositions of $p$-eigenstates, i.e. $\ket{0} \propto \int dp \;(\sum_{t=-\infty}^{\infty} S_p^t) \ket{p}  \propto \sum_{s=-\infty}^{\infty} \ket{p=s\sqrt{\pi}}$ while $\ket{1} \propto  \int dp \;e^{-i \sqrt{\pi} p}(\sum_{t=-\infty}^{\infty} S_p^t)  \ket{p}  \propto \sum_{s=-\infty}^{\infty} (-1)^s \ket{p=s\sqrt{\pi}}$ (using that $\sum_{t=-\infty}^{\infty} e^{-i 2 \sqrt{\pi}p t} \propto \sum_{s=-\infty}^{\infty} \delta(p=\sqrt{\pi} s)$).
Thus, both $\ket{0}$ and $\ket{1}$ have nonzero amplitude at integer multiples of $\sqrt{\pi}$ in $p$-space, but due to alternating phases, these amplitudes destructively interfere at odd multiples of $\sqrt{\pi}$ in the state $\ket{+}=\frac{1}{\sqrt{2}}(\ket{0}+\ket{1})$  \footnote{Please note the somewhat confusing $\sqrt{2}$ conversion factor between displacements and quadrature values, i.e. $q=\sqrt{2}{\rm Re}(\alpha)$ and $p=\sqrt{2}{\rm Im}(\alpha)$: the spacing between peaks in the code states in ${\rm Re}(\alpha)$ is $\sqrt{2\pi}$ while it is $2\sqrt{\pi}$ in $q$-space}. The operator $X \colon \ket{0} \leftrightarrow \ket{1}$ is given by $X=e^{-i \sqrt{\pi} p}$, as it shifts each eigenstate $\ket{q}$ by $\sqrt{\pi}$. Note that $X^2=S_p$ which equals $I$ only on the code space ($+1$ eigenspace of $S_p$) (see discussion in Appendix \ref{sec:phase}). One can write
\begin{align}
\ket{+} &=  \frac{1}{\sqrt{2}}(\ket{0}+\ket{1}) \propto \sum_{t=-\infty}^{\infty}S_q^t \ket{p=0}\notag \\
\ket{-}  &=  \frac{1}{\sqrt{2}}(\ket{0}-\ket{1})\propto \sum_{t=-\infty}^{\infty}S_q^t \ket{p=\sqrt{\pi}}, 
\end{align}
with $Z\colon  \ket{+} \leftrightarrow \ket{-}$ given by $Z=e^{i \sqrt{\pi}q}$ and $Z^2=S_q$. 

A simple way to understand the ideal preparation of a code state $\ket{0}$ (or $\ket{1}$) is through Eq.~(\ref{eq:perfect}): one starts with a $+1$ eigenstate of $Z$, namely $\ket{q=0}$ to which one applies $\Pi_{S_p=1}=\sum_{t=-\infty}^{\infty}S_p^t$ which is the projector onto the $+1$ eigenspace of $S_p$ \footnote{One can verify this by noting that $\Pi_{S_p=1}^2=\Pi_{S_p=1}$ and using the Poisson summation formula $\sum_{t=-\infty}^{\infty} e^{i \theta t} \propto \sum_s \delta(\theta=2 \pi s)$.}. 

\subsubsection{Approximate GKP Code States}

Naturally, Ref.~\cite{GKP} realized that the perfect code states in Eq.~(\ref{eq:perfect}) are unphysical as their preparation would require infinite squeezing and an infinitely sharp projection onto the $+1$ eigenspace of $S_p$. Ref. \cite{GKP} suggested using approximate code states: we can understand how this approximation comes about in, say, the definition of $\ket{0}$, as follows.

As it would take an infinite amount of squeezing to prepare $\ket{q=0}$ we replace this state by a finitely squeezed state (in $q$) centered around $q=0$, that is, a squeezed vacuum state, Eq.~(\ref{eq:sqvac}). Then we need to implement an approximate projection onto the $S_p=1$ eigenspace. Such projection could come about by approximately estimating the eigenvalue of the operator $S_p$ and then {\em post-selecting} the measurement outcome on this eigenvalue being $+1$ (while discarding other results). Such approximate projection by post-selection, using an ancilla qubit, has been considered in \cite{TM:GKPprep}, without making reference to any particular technology. However, any post-selection scheme will have a low probability of success and is in fact unnecessary. 
If one can implement a highly accurate phase estimation measurement of the unitary displacement operator $S_p$, estimating its eigenvalue as some $e^{i \theta}$, then one can also correct for such an eigenvalue shift
and shift, or displace back to the $+1$ eigenvalue code space. 

\begin{figure}[htb]
    \centering
    \subfigure[][]{
    \includegraphics{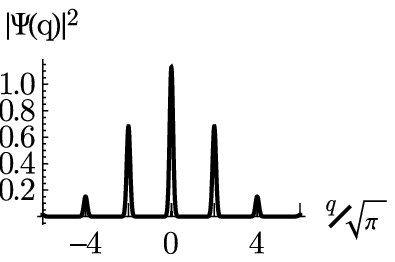}
    }
    \subfigure[][]{
    \includegraphics{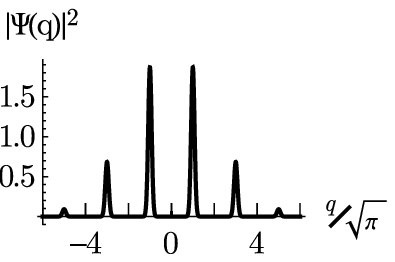}
    }
    \caption{%DW RESUBMIT CHANGE: removed color, "color online".
    Absolute value of the wave function of the approximate code state $\ket{0}_{\rm approx}$ (a) and $\ket{1}_{\rm approx}$ (b) for $\Delta=0.2$.}
\label{fig:code state}
\end{figure}

\begin{figure}[htb]
    \centering
    \subfigure[][]{
    \includegraphics[width=.45\hsize]{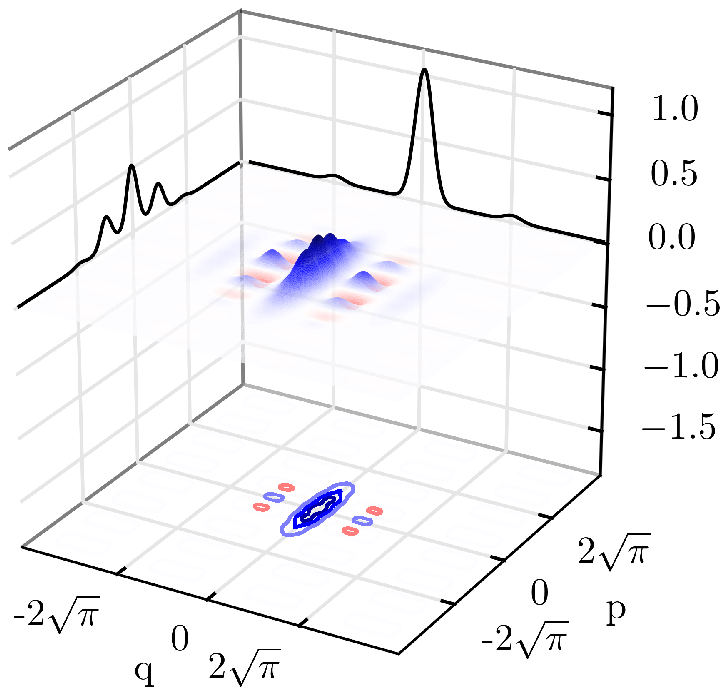} 
    }
    \subfigure[][]{
     \includegraphics[width=.45\hsize]{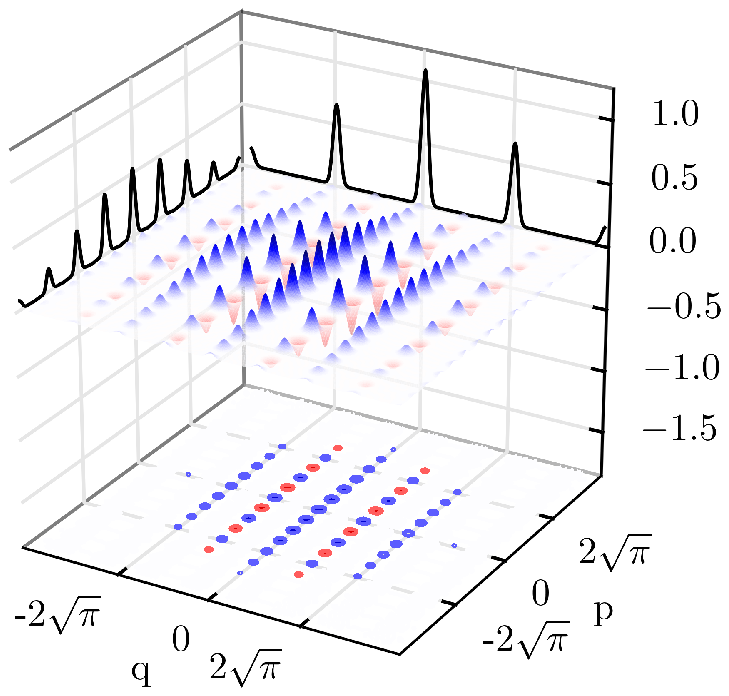} 
     }
    \caption{%DW RESUBMIT CHANGE: changed labels in the text, added description of types of plots shown. QUESTION: It is not necxessary to keep the reference to fig1, as the integral is also shown here, should we keep it?
    (Color online) Surface plot, contour plot and integral over $p, q$ of the Wigner function $W(p,q)$ for code states $\ket{0}_{\rm approx}$ with $\Delta=0.5$ (a) and $\Delta=0.2$ (b). For $q$ being an odd multiple of $\sqrt{\pi}$, the Wigner function rapidly oscillates between positive and negative values when $p$ is an even resp. odd multiple of $\sqrt{\pi}$. Integrating over the $p$-variable thus leaves, due to destructive interference, little amplitude for $q$ at odd multiples of $\sqrt{\pi}$, as can be seen in Fig.~\ref{fig:code state}.}
\label{fig:code state_wigner}
\end{figure}

 The formal definition of the approximate states is as follows. One replaces each delta-function in position $q$ by a squeezed Gaussian state while the uniform superposition over these localized states is replaced by a Gaussian envelope. For example, for the approximate $\ket{0}$ state, one starts with the squeezed vacuum state $\ket{\rm sq. vac}$ with squeezing parameter $\Delta^2=e^{-2r}$ 
\begin{equation}
\ket{\rm sq. vac}=\int \frac{dq}{(\pi \Delta^2)^{1/4}}e^{-q^2/(2\Delta^2)}\ket{q},  
\label{eq:sqvac}
\end{equation}
to which one applies a sum of displacements, as in Eq.~(\ref{eq:perfect}), with a Gaussian filter, i.e.
\begin{align}
\ket{0}_{\rm approx} &\propto \sum_{t=-\infty}^{\infty}e^{-2 \pi \tilde{\Delta}^2 t^2}D(t \sqrt{2\pi}) \ket{\rm sq. vac} \notag\\
&= \sum_{t=-\infty}^{\infty}\int e^{-2 \pi \tilde{\Delta}^2 t^2} e^{-(q - 2 t \sqrt{\pi})^2/(2 \Delta^2)} \ket{q} \text{d} q.\notag \\
\ket{\overline{1}}_{\rm approx} &\propto \sum_{t=-\infty}^{\infty}\int e^{-\pi \tilde{\Delta}^2 (2t+1)^2/2} e^{-(q - (2 t+1) \sqrt{\pi})^2/(2 \Delta^2)} \ket{q} \text{d} q.
\label{eq:approx}
\end{align}
%DW COMMENT: The equation above is broken, not sure how it can be fixed nicely. Leave it up to the editor?
The code state is thus a Gaussian-weighted sum  of displaced squeezed vacuum states.
 This code state and its Wigner function $W(p,q) = 1/\pi \int_{-\infty}^\infty e^{2ipx}\Psi^*(q+x)\Psi(q-x)\text{d} x$, where $\Psi(q)$ is the wave function of the state $\ket{0}_{\text{approx}}$ in the position basis, are depicted in Figs.~\ref{fig:code state} and  \ref{fig:code state_wigner} for $\tilde{\Delta}=\Delta=0.2$ which corresponds to 8.3 dB.
 %DW COMMENT: Not quite correct, we also do 0.5 in fig2. But I think it is ok as is, as it is simpler.
  Here and elsewhere in the paper we calculate dB as $10\log_{10}(G)=10 \log_{10} (\cosh^2(r)))$ where $G$ is the amount of gain through the amplifier and $r=\log (1/\Delta)$ \footnote{In some references squeezing dB is estimated as $10 \log_{10}(e^{2r})=-10 \log_{10}( 2 (\Delta q_{\rm sq})^2)$ where $(\Delta q_{\rm sq})^2$ is the variance of the squeezed state. For finite amounts of squeezing the difference between these estimates can be considerable, i.e. for $\Delta=0.2$ one would obtain $10 \log_{10}(1/\Delta^2) {\rm dB} \approx 13.98\, {\rm dB}$ of squeezing.}.
 
In the limit of $\Delta \rightarrow 0$ (infinite squeezing) where the width of the Gaussian envelope becomes arbitrarily broad, one obtains the perfect code states. It is worth noting that it is not essential that the parameters $\tilde{\Delta}$ and $\Delta$ are identical, nor that the filter is Gaussian: in \cite{BKP:protect} a very general form of approximate code states depending on two different filter functions was formulated.  

One can obtain the form of the code states in $p$-space by taking a Fourier transform. For example, the state $\ket{+}=\frac{1}{\sqrt{2}}(\ket{0}_{\rm approx}+\ket{1}_{\rm approx})$ has the following form when 
$\Delta/\sqrt{\pi}, \tilde{\Delta}\sqrt{\pi} \ll 1$, see \cite{GKP}:
\begin{align}
\ket{+}_{\rm approx} &\appropto  \sum_{t=-\infty}^{\infty}\int e^{-\Delta^2 p^2/2} e^{-(p - 2 t \sqrt{\pi})^2/(2 \tilde{\Delta}^2)} \ket{p} \text{d} p \notag \\
&\approx \sum_{t=-\infty}^{\infty}\int e^{-2 \pi \Delta^2 t^2} e^{-(p - 2 t \sqrt{\pi})^2/(2 \tilde{\Delta}^2)} \ket{p} \text{d} p,
\label{eq:approx_p}
\end{align}
where one observes that the roles of $\Delta$ and $\tilde{\Delta}$ are interchanged.
If one encodes in a bosonic mode, it is natural to choose similar approximations in $p$ and $q$ as free evolution of the state evolves these quadratures into each other.
This means that the choice $\Delta=\tilde{\Delta}$ is natural. The filter is Gaussian, so that we can use squeezed states which are Gaussian in their quadrature spread. 
It can be seen in Eqs.~(\ref{eq:approx}) and (\ref{eq:approx_p}) that the wave functions of approximate code states are even in $q$ and $p$, respectively. As the photon parity operator $P=e^{i\pi a^\dag a}$ transforms $q\to-q,\ p\to-p$, these code states will always have an even number of photons (this is not true for the approximate shifted code states in Eq.~(\ref{eq:approx_shift})).

For an approximate code state with parameter $\Delta$ one can explicitly calculate the mean number of photons $\overline{n}=\langle a^{\dagger} a\rangle$ and its variance $\sigma^2(n)=\langle (\overline{n}-a^{\dagger} a)^2 \rangle$.
Using that $\overline{n}= \langle\frac{1}{2}(p^2+q^2)+1\rangle = \langle q^2 \rangle +1$, $\Delta\ll 1/\Delta$ and approximating the infinite sum over $t$ by an integral, one obtains
$\overline{n} \approx \frac{1}{2\Delta^2}$. A similar calculation yields 
\begin{equation}
\sigma(n)  \approx \frac{1}{2\Delta^2} \sim \overline{n},
\label{eq:dev} 
\end{equation}
to  leading order in $1/\Delta$. This shows that the approximate code states are highly non-classical states with large fluctuations in photon number, i.e. scaling with the total number of photons while the approximation parameter $\Delta$ controls the number of photons in the state. We will return to the number of photons in the approximately-prepared code states in Section \ref{sec:photon}.  

%All states in the code space are examples of states which have a high sensitivity to phase-space displacements as was argued for so-called compass states in \cite{zurek:sub}.

\subsection{Previous Work}
\label{sec:prev}
The preparation of the code states has been discussed in several papers see e.g. \cite{GKP, TM:GKPprep, GK:osc, VSG:all_opt, PMVT:prep}.
In \cite{GKP} it was noted that an interaction $H_{\rm int}=q b^{\dagger} b$ where $q$ is the position operator of the oscillator (e.g. a cooled micromechanical system) in which the code state is to be encoded and $b$ is the annihilation operator of a different bosonic mode, would be useful in preparing the code states.  The protocols discussed in \cite{TM:GKPprep, VSG:all_opt} use post-selection in preparing the code states making the success probability extremely small, while we find that (almost no) post-selection is required by understanding the preparation step as phase estimation (see details in Section \ref{sec:num}). 

In \cite{mira_etal} the use of cat states $\ket{0}\propto \ket{\alpha}+\ket{-\alpha}$ and $\ket{1} \propto \ket{i\alpha}+\ket{-i \alpha}$ was explored for encoding a qubit in a cavity mode and dissipative dynamics was considered to preserve the even photon number code space.
Even though dissipative dynamics can drive a state to this code space, it does not imply that such dissipative dynamics realizes quantum error correction as it has to approximately drive an encoded state in this subspace to itself. Unlike the approximate GKP code, the cat state code by itself (implemented in circuit-QED in \cite{sun+:cat, vlastakis+:cat100}) is not fully protected against loss of photons. The detection of a photon number change by 1 constitutes a reversible change (which can be undone), but when one does not detect this change, the cat states decay irreversibly, nor is the code protected against the simultaneous loss of two photons.

In \cite{kirchmair+:kerr} multi-component cat states were dynamically generated in a cavity-mode using the Kerr nonlinearity of the cavity (see e.g. the description of such a protocol in \cite{book:haroche}). A multi-component cat state is a superposition of coherent states spaced at equal distances on {\em a circle} in phase space while the approximate code states are created by translations {\em along a line}.
It seems hard to define two approximately orthogonal qubit states using these multi-component cat states such that (1) the phase space amplitude peaks for $\ket{0}, \ket{1}$ are sufficiently separated (so as to protect against small shift errors) {\em and} (2) $\ket{+}$ and $\ket{-}$ are similarly superpositions of sufficiently separated peaks. 

\subsection{Quantum Error Correction and Protection}
\label{sec:shift}

The primary reason for defining the two-dimensional code space as the $+1$ eigenspace of $S_p$ and $S_q$ is that the code states are protected against small shift errors in phase space, of the form $e^{-i u p}$ and $e^{-i v q}$ with $|u|, |v| < \sqrt{\pi}/2$. Such shifts move the peaks of the code state in Fig.~\ref{fig:code state} as $e^{-i u p} \ket{q}=\ket{q+u}$ and $e^{-i v q} \ket{p}=\ket{p-v}$.
Thus when the shift error is less than half the shift represented by the operators $X$ or $Z$ in magnitude, one can undo the error by shifting back the state by the minimal amount.

Imagine that a shift error $e^{-i v q}$ has occurred and the eigenvalue of the check operator $S_p$ is perfectly measured by some means. One will estimate the eigenvalue of $S_p$ as $e^{i 2 \sqrt{\pi} v}$  (as $S_p e^{-i v q}\ket{\psi}=e^{i 2\sqrt{\pi}v} e^{-iv q} \ket{\psi}$ for a state $\psi$ in the code subspace, as follows from the commutation between $e^{-i v q}$ and $S_p$). For $|v|  < \sqrt{\pi}/2$, the phase $\theta=2 \sqrt{\pi} v\in (-\pi,\pi)$ is uniquely given and thus one can correct the shift error by learning the phase $\theta$ (and shifting back). For larger $|v|$, $\theta$ is consistent with two values of $v$ and choosing the wrong one leads to a logical qubit error.   Error correction for shift errors $e^{-i u p}$ works similarly by measuring the eigenvalue of $S_q$ (which becomes $e^{i 2 \sqrt{\pi} u}$).
In the phase estimation protocols that we propose, the goal is to approximately measure the eigenvalues of the unitary operator $S_p$ (and $S_q$). Once one knows the approximate eigenvalue of, say, $S_p$, one could in principle apply an appropriate corrective displacement such that the eigenvalue of $S_p$ equals 1.  In Appendix \ref{sec:phase} we argue that it is not necessary to do so as any subspace characterized by fixed eigenvalues of $S_p$ and $S_q$ is a good code space (one can compare this Phase frame to a Pauli frame used in the description of stabilizer codes). 

Perfect code states which are eigenstates of $S_p$ and $S_q$, can be parametrized \cite{GK:osc} as basis vectors $\ket{u,v}$ (which form a complete orthonormal basis for the oscillator space) defined as
\begin{align}
&\ket{u,v}=e^{-i u p} e^{-i v q} \ket{0},\notag \\
&u\in (-\sqrt{\pi},\sqrt{\pi}],\ v\in (-\sqrt{\pi/2},\sqrt{\pi/2}].
\end{align}
Note that one has $S_p \ket{u,v}=e^{i 2 \sqrt{\pi} v} \ket{u,v}$ and $Z \ket{u,v}=e^{i \sqrt{\pi} u} \ket{u,v}$. One can observe that a state $\ket{u,v}$ in $q$-space is simply a sum of peaks where the location of the delta-functions is shifted by $u$, while the wavefunction at each peak obtains a complex phase given by $\exp(-i v q)$.
One can similarly define approximate shifted code states as 
\begin{align}
&\ket{u,v, \Delta} \equiv e^{-i u p} e^{-i v q} \ket{0}_{\rm approx} \notag\\
&\propto \sum_{t=-\infty}^{\infty}\int e^{-2 \pi \Delta^2 t^2} e^{-(q -2 t \sqrt{\pi})^2/(2 \Delta^2)} e^{-i v q} \ket{q+u} \text{d} q
\label{eq:approx_shift}
\end{align}
For such an approximate shifted code state the Gaussian envelope is identical, but each peak within the Gaussian envelope has obtained a phasefactor $e^{-i v q}$ (we omit an overall phase factor) and the peaks are located at the shifted positions $q=u \mod 2\sqrt{\pi}$. Of course, the width of the Gaussian envelope with standard deviation $\sigma=1/\Delta$ should be larger than the maximal shift, i.e.  $\Delta \ll 1/\sqrt{\pi} \approx 0.56$, so that all shifts lead to approximate code states within the same overall envelope. It is the goal of the phase estimation protocols in this paper to produce such approximate shifted code states $\ket{u,v,\Delta}$.\\

It has been shown in \cite{GK:osc} that it is possible to do quantum error correction using perfect code states which have undergone some (coherent or stochastic) distribution of shift errors of strength $|u|,|v| < \sqrt{\pi}/6$, assuming otherwise perfect linear optical circuits. In Appendix \ref{sec:steane_qec} we review this way of doing quantum error correction. We will use the $\sqrt{\pi}/6$ value, corresponding to a phase uncertainty $\delta \theta$ of the stabilizer checks of at most $\pi/3$ (but a phase uncertainty of $\pi/6$ for the measurement of $X$ or $Z$), as a rough figure of merit to represent how well one can prepare a code state using phase estimation. We will refer to this threshold as the $\sqrt{\pi}/6$ {\em shift error threshold}. The `Steane quantum error correction' analyzed by Glancy and Knill already assumes the existence of an approximately-prepared code state for which we devise a phase estimation method in this paper. Through approximate phase estimation one can perform quantum error corection, hence the Glancy-Knill method may not be necessary to use at all.  

More formally, we define an {\em effective shift error rate} $P_{\rm error}^{\sqrt{\pi}/6}$ of a state which makes reference to the $\sqrt{\pi}/6$ threshold as follows.  Any density matrix $\rho$, i.e. an approximately prepared code state, can be written in the basis $\ket{u,v}$ \cite{GK:osc}, as
\begin{eqnarray}
\rho& = & \int_{-\sqrt{\pi}}^{\sqrt{\pi}} du \int_{-\sqrt{\pi}}^{\sqrt{\pi}} du' \nonumber \\
  & & \int_{-\sqrt{\pi}/2}^{\sqrt{\pi}/2} dv \int_{-\sqrt{\pi}/2}^{\sqrt{\pi}/2} dv' \rho_{uv,u'v'}  
\ket{u,v} \bra{u',v'}.
\label{eq:rho_shift}
\end{eqnarray}
The effective shift error rate $P_{\rm error}^{\sqrt{\pi}/6}$ is defined as
\begin{equation}
P_{\rm error}^{\sqrt{\pi}/6}= 1- \int_{-\sqrt{\pi}/6}^{\sqrt{\pi}/6} du \int_{-\sqrt{\pi}/6}^{\sqrt{\pi}/6} dv \; \rho_{uv,uv},
\label{eq:error}
\end{equation}
so that for a pure state $\ket{\psi}=\int_{-\sqrt{\pi}}^{\sqrt{\pi}} du \int_{-\sqrt{\pi}/2}^{\sqrt{\pi}/2} dv\; c(u,v) \ket{u,v}$, one has $P_{\rm error}^{\sqrt{\pi}/6}=1- \int_{-\sqrt{\pi}/6}^{\sqrt{\pi}/6} du \int_{-\sqrt{\pi}/6}^{\sqrt{\pi}/6} dv \; |c(u,v)|^2$.  For the approximate code state in Eq.~(\ref{eq:approx}) \cite{GK:osc} has shown that with $\overline{n} \approx 22$ photons, $\Delta=0.15$, the probability for shift errors beyond the $\sqrt{\pi}/6$ shift error threshold is at most $1\%$. 
If the shift errors are due to independent processes and the probability distribution $\rho_{uv, uv}=\rho_u \tilde{\rho}_v$ then one  
can use the effective shift error rate in $p$ or $q$ alone, i.e. $P_{\rm error,q}^{\sqrt{\pi}/6}= 1- \int_{-\sqrt{\pi}/6}^{\sqrt{\pi}/6} du \,\rho_u$ and $P_{\rm error,p}^{\sqrt{\pi}/6}= 1- \int_{-\sqrt{\pi}/6}^{\sqrt{\pi}/6} dv \; \tilde{\rho_v}$ to estimate the total error $P_{\rm error}^{\sqrt{\pi}/6}$.

Note that this error rate is only a rough figure of merit as it makes reference to the shift error threshold of $\sqrt{\pi}/6$ which has no intrinsic meaning if quantum error correction is performed through other means. For an approximate code state one can also estimate the $X$ or $Z$ error probability. The $X$ error probability is the probability that the state $\ket{0}_{\rm approx}$ is incorrectly identified as encoding $\ket{1}$ by means of {\em perfect} quantum error correction.
This is equal to the probability $P_{\rm error,q}^{\sqrt{\pi}/2}$ as shifts  with $|u| \geq \sqrt{\pi}/2$ lead to the state being incorrectly identified. In \cite{GKP} this error probability was approximately upperbounded by $\frac{2\Delta}{\pi} \exp(-\pi/4\Delta^2)$ which leads to a $1\%$ error probability for $\Delta=0.5$. Naturally, this error probability is more optimistic than the $\sqrt{\pi}/6$ threshold value, but since we expect the presence of additional (shift) errors during all protocols, using the threshold value seems a reasonable figure of merit. 

In practice, the code states are realized through an approximate noisy implementation of the phase estimation measurement of $S_p$ and $S_q$. Phase estimation by finite means, see Section \ref{sec:prep}, will typically output an estimate $\tilde{\theta}$ for $\theta$ such that ${\rm Prob}(|\tilde{\theta}-\theta| < \delta) > 1-\epsilon$ for some $\epsilon$ and $\delta$. This means that with probability at least $1-\epsilon$, one has $|\tilde{\theta}-\theta|=\delta \theta < \delta$ phase uncertainty, which in turn corresponds to $\delta v= \frac{\delta \theta}{2\sqrt{\pi}} < \delta/(2\sqrt{\pi})$ shift uncertainty or error with probability at least $1-\epsilon$. Note that when one measures the eigenvalue of an operator such as $Z=e^{i \sqrt{\pi}q}$ with phase uncertainty $\delta \theta <\delta$ (with high probability), then this corresponds to having a shift errror of strength at most $\delta u < \delta/\sqrt{\pi}$.

\subsection{Physical Sources of Errors as Shift Errors}
\label{sec:phys_noise}

In this section we discuss how levels of decoherence or inaccuracy {\em of an arbitrary nature} on the oscillator can be expanded into shift errors in $p$ and $q$ of small strength, at least when this noise is acting on states with a sufficiently low numbers of photons.  We will consider this for natural physical sources of noise and inaccuracy such as dephasing, photon loss and quartic (self-Kerr) interactions of the bosonic mode.

Following \cite{GKP}, any operator $E$ acting on a single bosonic mode can be expanded as $E= \int_{-\infty}^{\infty} d\gamma\;c(\gamma) D(\gamma)$ with complex coefficients $c(\gamma)$, i.e. an expansion into linear combinations of translations $D(\gamma)$ with complex $\gamma$ in phase space. One can formally obtain the coefficients $c(\gamma)$ as $c(\gamma)=\frac{1}{\pi}{\rm Tr}(E D(-\gamma))=\frac{1}{\pi^2} \int d\alpha \bra{\alpha} E D(-\gamma) \ket{\alpha}=\frac{1}{\pi}C_W^E(-\gamma)$ where $C_W^E(\lambda)$ is the symmetrically-ordered characteristic function of the operator $E$ \cite{book:haroche, book:mandel_wolf}, whose Fourier transform is the Wigner function of the operator $E$. Here $d\alpha= d{\rm Re}(\alpha)d{\rm Im}(\alpha)$. Thus in principle one can find an error expansion for any operator by evaluating $c(\gamma)$ and considering whether its support is concentrated on small values of $\gamma$. For example, the identity operator $I$ has $c(\gamma)=\delta^2(\gamma)$, a delta-function in the real and imaginary part of $\gamma$.

At a more intuitive level, one can understand however that the extent to which an expansion into small shifts is warranted, should depend on the number of photons in the state. Consider the action of an undesired rotation $\exp(-i\delta a^{\dagger} a)$ (or photon decay operator $\exp(-\delta a^{\dagger} a)$) in phase space. For simplicity, we apply it to a coherent state $\ket{\alpha}$: it is clear that such rotation corresponds to a larger state-dependent translation for larger $\alpha$, i.e. $\exp(-i\delta a^{\dagger} a) \ket{\alpha}=\ket{\alpha \exp(-i\delta)} \propto D(\alpha(\exp(-i\delta)-1)) \ket{\alpha}\approx D(-i \alpha \delta)\ket{\alpha}$ for $\delta \ll 1$. If $\exp(-i\delta a^{\dagger} a)=\int d\gamma\; c(\gamma) D(\gamma)$ were only supported on $\gamma$ with $|\gamma|^2 \leq f(\delta)$ with some function $f(\delta)$ which is independent of the number of photons, then this contradicts the fact that there are states $\ket{\alpha}$ with $n=|\alpha|^2$ large enough to be undergoing a large displacement $|\alpha \delta| > |\gamma|$ \footnote{In \cite{GKP} the phase space rotation $\exp(-i \delta a^{\dagger} a)$ was expanded in shifts $\gamma$ such that $|\gamma|^2 \sim \delta$ (so   small shift errors when $\delta$ is small), but the stationary phase approximation that was invoked was not proved in any rigorous fashion.}.

This implies that one should consider expansions (of physical noise
operators in terms of shifts) which assume an upper bound on the number
of photons in the oscillator space: let us call this upper bound on the
number of photons $n_{\rm max}$. In that case it is clear from the above
example that one can only make an expansion in small shifts, of strength
at most $\sqrt{\pi}$, of $\exp(-i \delta a^{\dagger} a)$ when {\em at
least} one satisfies the inequality $|\delta| n_{\rm max}^{1/2} <
\sqrt{\pi}$. A similar argument can be given for a quartic interaction
of form $\exp(-i \epsilon (a^{\dagger} a)^2)$. For such quartic
interaction one can write, assuming $\epsilon \ll 1$,
\begin{align}
\bra{\alpha} \exp(i \epsilon &(a^{\dagger} a)^2) a \exp(-i \epsilon(a^{\dagger} a)^2) \ket{\alpha} \notag \\
&\approx \alpha \exp(-i |\alpha|^2\epsilon)) \exp(-|\alpha|^2 \epsilon^2/2).
\end{align}
Thus ignoring fluctuations, one has an average phase space rotation
(non-linear phase shift) in addition to coherent state amplitude decay.
As before, such average rotation and decay is only expressible in terms
of small shifts of strength less than $\sqrt{\pi}$ when at least one
obeys the inequality $|\alpha|^3 |\epsilon| \leq |\epsilon| n_{\rm
max}^{3/2} < \sqrt{\pi}$.

We note that these conditions are necessary, but not here proved to be
sufficient as we have only applied these operators to coherent states
$\ket{\alpha}$ (while for example a superposition of Fock states
$\ket{{\rm vac}}+\ket{n_{\rm max}}$ undergoes a rate of change
proportional to $\epsilon n_{\rm max}^2$ by $\exp(-i \epsilon
(a^{\dagger} a)^2)$).

Since the approximate code states are superpositions of displaced
squeezed vacua, we believe however that these criteria do give a good
indication of whether a small-shift expansion is warranted and also
capture the strength of shift errors. For the proposed protocol in
Section \ref{sec:prop} we will show that the expected process of photon
loss during the preparation of the code state has sufficiently small
error rate so as to be deeply into this small-shift error regime. We
will also discuss the (unwanted) cavity nonlinearities in the effective
transmon qubit cavity Hamiltonian given the physical strength of
parameters in a possible experiment.

\subsubsection{Shift Error Expansion}
\label{sec:phys_shift}

Here we describe a general method to expand an error operator $E$ in terms of shift errors.
 The error operator
$E$ can arise either from open system dynamics or an unwanted unitary
transformation. General open system dynamics of the bosonic mode can be
modeled by a Lindblad equation of the form
\begin{equation}
\dot{\rho}=-i [H_{\rm ideal}(t)+V(t), \rho]+{\cal D}(\sqrt{\kappa}a)(\rho)+{\cal D}(\sqrt{\gamma}a^{\dagger} a) (\rho),
\label{eq:opensys}
\end{equation}
with the compactly-defined superoperator ${\cal D}(X)(\rho)= X \rho
X^{\dagger}-\frac{1}{2}(X^{\dagger} X\rho+\rho X^{\dagger} X)$. Here
$H_{\rm ideal}(t)$ is some ideal (time-dependent) dynamics (the
execution of the phase estimation protocol of $S_p$ say, or some gate implementation) and $V(t)$ a possible perturbation or correction
(e.g. an unwanted self-Kerr term $K (a^{\dagger} a)^2$). The Lindblad
equation can model photon loss $D(\sqrt{\kappa}a)$ of rate $\kappa$ or
dephasing $D(\sqrt{\gamma} a^{\dagger}a)$ of rate $\gamma$.
For
simplicity, one can consider the effects of the different sources of
noise separately (and ignore $H_{\rm ideal}(t)+V(t)$) and construct a
superoperator for the evolution in a short time interval, as ${\cal
S}(\rho(t+\tau))=\sum_i A_i \rho(t) A_i^{\dagger}$. For the process of
photon loss one has
$A_0=I-\frac{\tau \kappa}{2} a^{\dagger} a+O((\tau \kappa)^2n_{\rm
max}^2)$ and $A_1=\sqrt{\kappa\tau}a$. For dephasing one has
$A_0=I-\frac{\tau \gamma}{2}(a^{\dagger} a)^2+O((\tau \gamma)^2 n_{\rm
max}^4)$ and $A_1=\sqrt{\gamma \tau} a^{\dagger} a$. These Kraus
operators are examples of error operators $E$.
It is clear that one can
express such operators as a low-order polynomials in $a$ and $a^{\dagger}$ and we
will describe how such a low-order polynomial can be expanded in shift errors.

Similarly, a unitary error operator $E$ can be an over-rotation $\exp(-i
\delta a^{\dagger} a)$ or a nonlinearity $\exp(-i \epsilon (a^{\dagger}
a)^2)$ where $\delta, \epsilon$ are given by an error strength times a
time-scale. Assuming that $\delta n_{\rm max} \ll 1$ or $\epsilon n_{\rm
max}^2 \ll 1$, one can Taylor expand this in terms of $I$ and low powers
of $a$ and $a^{\dagger}$.

We now write a simple Taylor expansion in terms of correctable and
uncorrectable shift errors for the operator $\sqrt{\delta} a \equiv x_1+
i x_2$ with
$x_1=p\sqrt{\delta/2}$ and $x_2=q\sqrt{\delta/2}$ assuming $\delta n_{\rm max} \ll 1$. This expansion and
its hermitian conjugate can then be used to expand operators such as
$a^2, a^{\dagger}a$ etc. With the Taylor expansion of $\arcsin(x)
\approx x+\frac{x^3}{6}+\frac{3 x^5}{40}+\frac{5 x^7}{112}+O(x^9)$, we
can expand $x_1+i x_2$ in terms of $\sin(x_1),\sin^3(x_1)$ and
$\sin(x_2), \sin^3(x_2)$ etc. For example, in lowest-order one has
\begin{align}
\sqrt{\delta} a &=\sin (p\sqrt{\delta/2})+\sin^3(p\sqrt{\delta/2})/6 \notag \\
&+i(\sin(q\sqrt{\delta/2})+\sin^3(q\sqrt{\delta/2})/6)+O((\delta n_{\rm max})^{5/2}),
\label{eq:shift_expand}
\end{align}
where the $\sin(p\sqrt{\delta/2})$ and $\sin^3(p\sqrt{\delta/2})$ function (and similarly $\sin(q\sqrt{\delta/2})$) can be expanded in terms of 
$e^{\pm i p\sqrt{\delta/2}}$, $e^{\pm 2i \sqrt{\delta/2}}$, $e^{\pm 3i \sqrt{\delta/2}}$ etc.
In general
we can thus write $E=\sqrt{\delta} a=E_{\rm correctable}+ E_{\rm
uncorrectable}$ where $E_{\rm correctable}$ contains all terms in the
Taylor expansion up to odd $k$-th order such that $k\sqrt{\delta/2}
\approx \sqrt{\pi/2}$ and $||E_{\rm uncorrectable}|| \leq O((\delta
n_{\rm max})^{1+k/2})$.
We can use this to expand an operator
$(\sqrt{\delta} a)^p$ (or $(a^{\dagger}a)^{p/2}$ or $(a^{\dagger})^p$)
as $(\sqrt{\delta} a)^p=E_{\rm correctable}+ E_{\rm uncorrectable}$
where $|| E_{\rm uncorrectable}||=O((\delta n_{\rm max})^{1+k/2})$ with
$k\sqrt{\delta} \approx \sqrt{\pi}/p$.

This shows that on an encoded state the amplitude of an uncorrectable
error can be reduced {\em by several orders of magnitude}, depending on the
strength of the errors.

One thing to note is that in order to systematically expand noise
processes in terms of shift errors one should develop the full Kraus
error operator or undesired rotation to a certain order in $\delta
n_{\rm max}$. In other words, depending on which order in $\delta n_{\rm max}$ one chooses, one
includes higher-order terms in the expansion of the error Kraus
operators $A_i$ or the unitary.

One can interpret the protection that the code offers with the following
example. Assume that photon loss from a cavity occurs for some time $t$
at rate $\kappa$ such that $P\equiv \kappa t n_{\rm max} < 1$. Without
encoding, the error operator $E=\sqrt{\kappa t} a$ produces an
uncorrectable error with probability $\sim P$. For a cat-state encoding
(see Sec.\ref{sec:prev}) this process is correctable, but for the cat
encoding, a two-photon loss error operator $E=\kappa t a^2$ is
uncorrectable (as it is proportional to $Z$). Hence
uncorrectable errors happen with probability $\sim P^2$.

If we expand the 2-photon loss error operator in terms of correctable
and incorrectable shift errors, the probability for the uncorrectable
term is of order $(\kappa \tau n_{\rm max})^{1+k/2}=P^{1+k/2}$ where
$k\approx \frac{1}{2}\sqrt{\frac{\pi n_{\rm max}}{P}}$. Thus for small
$P\ll 1$ the probability for an uncorrectable error resulting from
2-photon loss can be much reduced if one uses a code that can correct
shift errors as compared to a code which can only correct single-photon
loss events.

\section{Phase Estimation}
\label{sec:prep}

The measurement of the eigenvalue $e^{i\theta}$ of a unitary operator $U$, and the simultaneous projection of the input state onto the corresponding eigenstate $U\ket{\psi_{\theta}}=e^{i\theta}\ket{\psi_{\theta}}$, is called phase estimation for $U$. In our case we have $U=S_p=D(\sqrt{2\pi})$, say. 
Phase estimation can be executed by repeatedly running a circuit of a general form depicted in Fig.~\ref{fig:SPE} for varying $k$ and phases $\varphi$. Many variants of phase estimation exist (see e.g. a recent analysis in \cite{SHF:PE}) depending how or whether one varies $k$, and/or whether one allows $\varphi$ to depend on earlier qubit measurements, so using feedback, and how one infers the phase from the information obtained from the sequence of qubit measurement outcomes.

We will first consider standard phase estimation as it is used in Shor's factoring algorithm and argue that this method does not give the kind of approximate code state that we are looking for, at least not when we wish to use low numbers of photons. The standard phase estimation has the same performance as Kitaev's phase estimation described in \cite{KSV:computation}; both require realizations of $U^{2^k}$ for increasing $k$.

The best form of approximate phase estimation optimizes the accuracy on the phase given a mean number of photons $\overline{n}$ in the approximately prepared code state as we always work with a bounded number of photons. In this respect it can be noted that for the standard phase estimation protocol described below, the phase uncertainty scales as $\delta \theta \sim 2^{-M} \sim \frac{1}{\sqrt{\overline{n}}}$, that is, shot-noise limited. Protocols which can achieve Heisenberg-limited scaling, that is $\delta \theta \sim \frac{1}{\overline{n}}$, have been considered and analyzed in detail in \cite{higgins+:nature, higgins+:heis, berry+:phase} (with an improved accuracy analysis in \cite{KLY:PE}). One scheme that can reach the Heisenberg-limit is a modified form of the Kitaev's phase estimation in which the circuit for each controlled-$D(2^k \sqrt{2\pi})$ in Fig.\ref{fig:SPE} is repeated a number of times depending on $k$ and $M$ \cite{berry+:phase}. One can thus expect that this Heisenberg-limited scheme performs optimally, but we do not consider it here as the experimental realization of the simpler schemes will already be demanding. When one allows for a larger number of ancilla qubit rounds, one can expect that switching to such scheme which uses $U^2$ or $U^4$ etc. is better.

In sections \ref{RPE} and \ref{sec:feedback} we analyze two simple phase estimation protocols, repeated or non-adaptive phase estimation and phase estimation by feedback or adaptive phase estimation. These schemes only use $U$ and thus no increasingly large displacements (microwave power).

\subsection{Standard Phase Estimation}

\begin{figure*}[h!tb]
\centering
\includegraphics{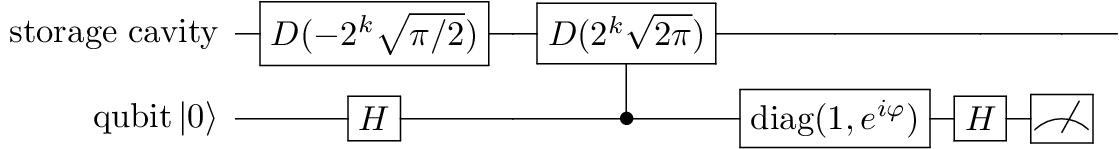}
%\centering
%\parbox{150pt}{
%\Qcircuit @C=1em @R=1em {
%\lstick{\mbox{storage cavity}} & \gate{D(-2^k\sqrt{\pi/2})} & \qw & \gate{D(2^k\sqrt{2\pi})} & \qw & \qw & \qw & \qw \\
%\lstick{\mbox{qubit} \ket{0}} & \gate{H} & \qw & \ctrl{-1} &  \gate{{\rm diag}(1,e^{i\varphi})} & \gate{H} & \meter
%}}
\caption{Phase estimation for the unitary operator $S_p=e^{-i 2\sqrt{\pi}p}=D(\sqrt{2 \pi})$ where $D(\alpha)$ is the displacement operator. Assume that phase estimation uses $M$ ancillas prepared in $\ket{00 \ldots 0}$. The circuit in this Figure is repeatedly executed for $k=M-1, \ldots, 0$ starting at $k=M-1$. The phase $\varphi$ in the single qubit rotation around the $z$-axis, ${\rm diag}(1,e^{i\varphi})$, will depend on the outcomes of {\em all} the previously measured ancillas. 
This sequential realization of phase estimation is identical to normal phase estimation as it merely uses a semi-classical realization of the quantum Fourier transform. Note that the circuit is identical to one in which ${\rm diag}(1,e^{i\varphi})$ is moved before the controlled-displacement gate, which is the form of the quantum circuit in \cite{higgins+:nature}.
Prior to the controlled-displacement gate, the cavity is (unconditionally) displaced so that the code states are symmetrically centered around the vacuum state and we minimize the total number of photons.}
\label{fig:SPE}
\end{figure*}

Any method for estimating the phase by sequentially acquiring bits of information is information-theoretically bounded: it is clearly optimal if each acquired bit gives us, with certainty, one additional bit of the binary expansion of $\frac{\theta}{2\pi}=0.\theta_1 \theta_2\ldots $, starting with the most significant bit $\theta_1$ etc. 
This is close to what the standard phase estimation protocol, as it is invoked in Shor's algorithm, achieves \cite{book:nielsen&chuang}.
It is known that this phase estimation protocol can be realized sequentially, see e.g. \cite{kitaev:stabilizer, ME:hidden, cleve+:revisited}, by coupling a sequence of $M$ single qubits to the input state and applying controlled-$U^{2^k}$ gates ($k=0,\ldots$) with the ancilla qubit as control. This can be understood by replacing the Fourier transform in the standard phase estimation by the semi-classical Fourier transform whose quantum circuit is one in which the first qubit is measured and its outcome is used to apply single qubit gates on the remaning $M-1$ qubits. The next qubit is measured and again its outcome determines single qubit gates on the remaining $M-2$ qubits etc., see Fig. 5.1 in \cite{book:nielsen&chuang}. In this way the whole procedure can be implemented sequentially, in rounds with one active control qubit per round.
We depict one round of such a phase estimation protocol for $U=S_p=D(\sqrt{2\pi})$ in Fig.~\ref{fig:SPE}.
We can consider the measurement operator $M_x$ that is applied to the input state, upon getting an $M$-bit outcome $x  \in\{0,1\}^M$, with $U=D(\sqrt{2\pi})$, including the compensating displacements in each circuit in Fig.~\ref{fig:SPE} which center the code state around the vacuum state. One obtains $M_x \ket{\psi_{\rm input}}$, with
\begin{align}
M_x &=\frac{1}{2^M} \sum_{t \in \{0,1\}^M} e^{-2 \pi i  x t /2^M} D( \sqrt{2\pi} t) D(-\sqrt{2\pi}\frac{2^M-1}{2}) \notag \\
&\appropto \sum_{t =-2^{M-1}}^{2^{M-1}} e^{-2 \pi i  x t /2^M} D( \sqrt{2\pi} t).
\label{eq:meas_pe}
\end{align}
(Here $\appropto$ relates to the fact that we simply approximate $(2^M-1)/2 \approx 2^{M-1}$). For phase estimation one can prove that, using $M=\tilde{M}+\log_2(\frac{1}{2}+\frac{1}{2 \epsilon})$ ancilla qubits, we obtain the best possible $\tilde{M}$-bit-estimate of the binary expansion of $\frac{\theta}{2\pi}$ with probability of success at least $1- \epsilon$ (when one chooses $\tilde{M}=M$, $P_{\rm success}=\frac{4}{\pi^2}$). Let $\tilde{\theta} \in (-\pi,\pi]$ represent this $\tilde{M}$-bit estimate.  Let $x_{\rm round}$ be the bitstring outcome $x$ rounded off to its $\tilde{M}$-most significant bits. For $x_{\rm round}/2^{\tilde{M}} \leq 1/2$, we choose $\tilde{\theta}=\frac{ 2\pi x_{\rm round}}{2^{\tilde{M}}}$, while for $x_{\rm round}/2^{\tilde{M}} > 1/2$, one takes $\tilde{\theta}=\frac{ 2\pi x_{\rm round}}{2^{\tilde{M}}}-2\pi$.

Knowing the first $\tilde{M}$ bits of $\theta/(2\pi)$ leads to an error $\delta \theta\leq 2\pi 2^{-\tilde{M}}$. This means that one can prepare an approximate $S_p$ eigenstate with $P_{\rm success} \geq \frac{9}{16}$ (for which $M=\tilde{M}+2$) such that the shift error is less than the $\sqrt{\pi}/6$ shift error threshold (correspond to $\pi/3$ phase error, see Section \ref{sec:shift}) if we take $M \geq 4$. 

There are a few ways to see that this method of measuring the eigenvalue of $S_p$ is not well suited for the approximation that we are seeking. The protocol requires the implementation of controlled-$D(2^k\sqrt{2\pi})$ for $k=0,\ldots M-1$, Fig.~\ref{fig:SPE}. This means that the number of photons in the cavity mode grows exponentially with the number of ancilla qubits used. More precisely, one can calculate the number of photons in an approximate code state produced by applying phase estimation of $S_p$ onto the squeezed vacuum state in Eq.~(\ref{eq:sqvac}), $\ket{\psi_x} \propto M_x \ket{\rm sq. vac}$. One has
\begin{align}
\overline{n}_x &=\frac{\bra{\rm sq. vac} M_x^{\dagger}a^{\dagger} a M_x \ket{\rm sq. vac}}{\bra{\rm sq. vac} M_x^{\dagger}  M_x \ket{\rm sq. vac}} \notag \\
&\approx 2 \pi \frac{1}{2^M}\sum_{k=-2^{M-1}}^{2^{M-1}} k^2 + \overline{n}_{\text{sq}} \notag \\
&=2\pi \left(\frac{2^{2M-2}}{3}+\frac{2^{M-1}}{2}+\frac{1}{6} \right)+\overline{n}_{\rm sq.},
\nonumber
\end{align}
where $\overline{n}_{\text{sq}}$ is the expected number of photons of the squeezed vacuum state $\ket{\rm sq. vac}$.
Here we have used that $D(-\alpha) a D(\alpha)=a+\alpha$ and $\bra{\rm sq. vac} D(\sqrt{2\pi}(t-t')) \ket{\rm sq. vac}\approx 0$ for $|t-t'| \geq 1$ (the numerical value is $O(10^{-13})$ for $\Delta = 0.2$).
In principle $\overline{n}_x$ depends on the outcome $x$, but the fluctuations with $x$ vanish when we approximate overlaps between different displaced squeezed states by 0. To leading order in $M$ we have $\overline{n}_x  \approx 2 \pi \frac{2^{2M}}{12}+\overline{n}_{\rm sq.}$, scaling exponentially with $M$ as expected. For $M=4$, one has $\overline{n} \approx 134+\overline{n}_{\rm sq}$. This number is very high compared to the 22 photons in a Gaussian-approximate code state with $\Delta=0.15$ which has high probability to be below the shift error threshold.

Another way to understand this is to consider the form of the measurement operator in Eq.~(\ref{eq:meas_pe}). Equating $x \approx x_{\rm round}$, we have $M_{x\rightarrow \tilde{\theta}} \appropto \sum_{t =-2^{M-1}}^{2^{M-1}} e^{- i  \tilde{\theta} t} D( \sqrt{2\pi} t)$. One can compare this measurement operator to the projector onto the space with $S_p=e^{i\theta}$ which equals
\begin{equation}
\Pi_{S_p=\exp( i \theta)}=\sum_{t=-\infty}^{\infty} e^{ -i \theta t} D(\sqrt{2\pi}t),
\end{equation}
and note the similarity between the two operators.
This shows that the filter used in standard phase estimation is a wide box-car filter with a hard cut-off, while we are seeking a smooth Gaussian filter. 
If we were to choose controlled-displacement operators with exponentially-growing displacements, one has to show that the duration of these gates is not necessarily exponentially-increasing as such growth in time would lead to more loss and decoherence during the QEC cycle and code state preparation.
In principle large controlled-displacements can be obtained by large microwave power, see Appendix \ref{sec:direct}, but one may expect that the inaccuracy and undesired side-effects of exponentially-large displacements are also exponentially increasing.

\subsection{Phase Estimation by Repetition (Nonadaptive)}
\label{RPE}

\begin{figure*}[h!tb]
\centering
%DW COMMENT: this is not quite aligned, but I think fixing for PRA is an unnecessary change, as it will probably be done by the editor.
\includegraphics{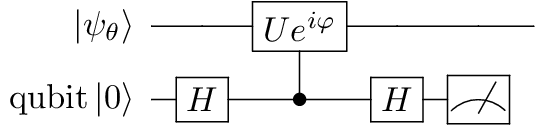} $=$ \includegraphics{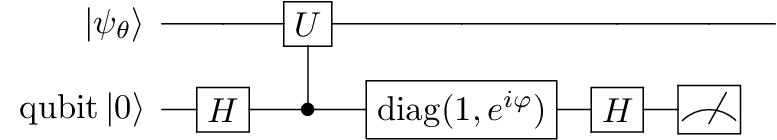}

%\parbox{150pt}{
%\Qcircuit @C=1em @R=1em {
%\lstick{\ket{\psi_{\theta}}} & \qw & \gate{Ue^{i\varphi}} & \qw & \qw & \qw \\
%\lstick{\mbox{qubit} \ket{0}} & \gate{H} & \ctrl{-1} & \gate{H} & \meter
%}
%}
% \makebox[60pt][c]{=}
%\parbox{150pt}{
%\Qcircuit @C=1em @R=1em {
%\lstick{} & \qw & \gate{U} & \qw & \qw & \qw & \qw \\
%\lstick{\ket{0}} & \gate{H} & \ctrl{-1} & \gate{{\rm diag}(1,e^{i\varphi})} & \gate{H} & \meter
%}
%}
\caption{
Phase estimation for a unitary operator $U$, with varying (feedback) phase $\varphi$, and $U\ket{\psi_{\theta}}=e^{ i \theta}\ket{\psi_{\theta}}$.  For repeated phase estimation one chooses either $\varphi=0$ and $\pi/2$ for half the number of rounds.
In a simple form of phase estimation with feedback the phase $\varphi$ is changed after each outcome so as to optimize the sensitivity of the probability distribution with respect to the currently estimated value for $\theta$. }
\label{fig:ape}
\end{figure*}

The simplest way to estimate $\theta \in (-\pi,\pi]$ of a unitary operator $U\ket{\psi_{\theta}}=e^{i\theta} \ket{\psi_{\theta}}$ is to repeat the quantum circuit in Fig.~\ref{fig:ape} with $\varphi=0$ and $\varphi=\pi/2$. For the $\varphi=0$ measurement, each ancilla qubit then has a probability for outcome 0 equal to $P_{\varphi=0}(0|\theta)=\frac{1}{2}(1+\cos(\theta))$ while for the $\varphi=\pi/2$ measurement one has $P_{\varphi=\pi/2}(0|\theta)=\frac{1}{2}(1-\sin(\theta))$. Note that a simple repetition of the $\varphi=0$ measurement is insufficient since $P_{\varphi=0}(0|\theta)$ is the same for $\theta$ and $-\theta$. Thus one chooses $\varphi=0$ for half of the number of rounds/ancilla qubits and $\varphi=\pi/2$ for the other half \cite{kitaev:stabilizer}.

In an adaptive phase estimation, see Section \ref{sec:feedback}, one takes into account that the sensitivity of the probability distribution $P_{\varphi}(0|\theta)=\frac{1}{2}(1+\cos(\theta+\varphi))$ to $\theta$, i.e. $\frac{d P_{\varphi}(0 |\theta)}{d\theta}$, is a function of $\theta$.
One would like to optimize this sensitivity by choosing values for {\em the feedback} phase $\varphi$ which depend on previous measurement outcomes. In this optimization one chooses a next phase $\varphi$ such that the measurement with that $\varphi$ maximizes the sharpness of the resulting inferred probability distribution over $\theta$. This method has been analyzed and described in detail in \cite{BWB01, berry+:phase}. \\

\begin{figure*}[htb]
\centering
\includegraphics{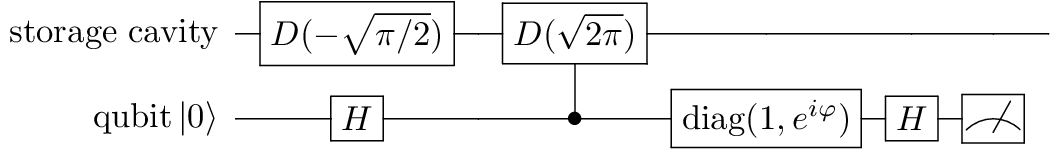}
%\parbox{150pt}{
%\Qcircuit @C=1em @R=1em {
%\lstick{\mbox{storage cavity}} & \gate{D(-\sqrt{\pi/2})} & \qw & \gate{D(\sqrt{2\pi})} & \qw & \qw & \qw & \qw \\
%\lstick{\mbox{qubit} \ket{0}} & \gate{H} & \qw & \ctrl{-1} &  \gate{{\rm diag}(1,e^{i\varphi})} & \gate{H} & \meter
%}}
\caption{One round of phase estimation for the unitary operator $S_p=e^{-i 2p \sqrt{\pi}}=D(\sqrt{2 \pi})$ where $D(\alpha)$ is the displacement operator. The phase $e^{i\varphi}$ of the $R_z$-qubit rotation can be either fixed to $\varphi=0$ and $\pi/2$ or adaptively changed per round depending on previous outcomes as in phase estimation by feedback, Section \ref{sec:feedback}. Prior to the controlled-displacement gate, the cavity is (unconditionally) displaced so that the code states are symmetrically centered around the vacuum state and we minimize the total number of photons.}
\label{fig:rpe}
\end{figure*}

One round of phase estimation for $U=S_p$ is depicted in Fig.~\ref{fig:rpe}. Instead of starting with an eigenstate, one would like to understand how one approximately projects onto an eigenstate of $S_p$ using this repeated measurement. We can first consider the action of this circuit on a coherent state $\ket{\alpha}$ when the qubit is measured as $x=0,1$:
\begin{equation}
\ket{\alpha} \rightarrow \ket{\alpha-\sqrt{\pi/2}}+(-1)^x e^{i \varphi} \ket{\alpha+\sqrt{\pi/2}},
\label{eq:displace}
\end{equation}
where the resulting state has not been normalized. Repetition of such a circuit on the resulting output state will thus produce a sum of coherent states on a line, each with a phase which depends on the feedback phase and the measurement outcome.
If all phases add constructively, this distribution of amplitudes/weights of coherent states will be as in a Pascal triangle, hence binomial, as each coherent state gets split in two equidistant coherent states at every round, see e.g. Fig.~\ref{fig:phasespace_sketch}(a) in Appendix \ref{sec:direct}.

A more formal way of showing that the filter of this protocol is binomial and thus approximately Gaussian is as follows.  Consider the measurement operator for $M$ rounds with $\varphi=0$. An outcome bitstring $x\in \{0,1\}^M$ will correspond to a measurement operator which can be labeled by the Hamming weight $k=w_H(x)$ as the order of the outcomes of $1$s and $0$s is irrelevant. The measurement operator equals
\begin{align}
&M_k \propto  (I+D(\sqrt{2 \pi}))^{M-k} (I-D(\sqrt{2 \pi}))^{k} D(- M\sqrt{\pi/2}) \notag \\
&= \sum_{p_1=0}^k \sum_{p_2=0}^{M-k}{k \choose p_1} {M-k \choose p_2} (-1)^{p_1} D(\sqrt{2\pi}(p_1+p_2-M/2)). \nonumber 
\end{align}
When all measurement outcomes $x_i=0$ (from which one would conclude that $\theta=0$) the measurement operator has the simple form of a binomial sum of displacements:
\begin{align}
M_{k=0} &\propto \sum_{m=0}^{M}{M \choose m} D(\sqrt{2\pi}(m-M/2)) \notag \\
&\appropto \sum_{t=-M/2}^{M/2}e^{-2 t^2/M} D(\sqrt{2\pi}t).\notag
\end{align}
One can roughly identify this measurement operator with the Gaussian-filtered projection operator onto approximate code state in Eq.~(\ref{eq:approx}) with $\tilde{\Delta}^2 \approx \frac{1}{\pi M}$ (note that the measurement operator inevitably has a hard cut-off while the Gaussian-filter does not). This shows that for $\tilde{\Delta}=0.2$ one can choose $M=25/\pi \approx 8$ with a much lower number of photons than in regular phase estimation.

Instead of postselecting on this outcome, we want to use the measurement data to estimate the value of $\theta$, and include the data from the $\varphi=\pi/2$ measurement. We can analyze the efficiency of this method using the Chernoff bound; this argument is explicitly developed in \cite{higgins+:heis}. Section \ref{sec:num} gives numerical details on the phase uncertainty and the number of photons. Assume that we use $M/2$ ancilla qubits for $\varphi=0$ and $M/2$ qubits for $\varphi=\pi/2$ such that $P_{\varphi=0}(0)$ is estimated as $\tilde{P}_{\varphi=0}$ and $P_{\varphi=\pi/2}(0)$ is estimated as $\tilde{P}_{\varphi=\pi/2}$. The estimate $\tilde{\theta}$ is chosen as
\begin{align}
&\tilde{\theta}=\arg (2\tilde{P}_{\varphi=0}-1-i(2\tilde{P}_{\varphi=\pi/2}-1)), \notag \\
&\tilde{\theta} \in (-\pi,\pi].
\label{eq:theta_est_sim} 
\end{align}
Using a Chernoff bound in both cases, ${\rm Prob}(|\tilde{P}-P| \geq \delta) \leq 2 e^{-2 \delta^2 M}$ and some further bounding arguments \cite{higgins+:heis}, one can show that 
${\rm Prob}(\delta \theta \geq \frac{\pi}{3} \sqrt{f(M)/M}) \leq 4 \exp(-3 f(M)/16)$ for any function $f(M) \leq M$ for $M\geq 1$. This implies that the probability to prepare a code state with phase uncertainty below $\pi/3$ is at least $1-4 \exp(-3 M/16)$.
This argument shows that the number of ancillas should be at least $M \geq 8$ and the probability of failure will then rapidly vanish. We can note that the phase uncertainty $\delta \theta \sim \frac{1}{\sqrt{M}}$ for $M$ rounds, each of which adds $O(1)$ photons to the state, hence scaling in the expected (shot-noise) way.

\subsection{Phase Estimation with Feedback (Adaptive)}
\label{sec:feedback}

The uncertainty of a probability distribution $P(\theta)$ over phases $\theta$ can be measured by the Holevo phase variance defined as $V[P(\theta))]=S[P({\theta})]^{-2}-1$ with the {\em sharpness} $S[P(\theta)]\in [0,1]$ given by
\begin{equation}
S[P({\theta})]\equiv |\langle e^{i\theta}\rangle |=\left|\int_{-\pi}^{\pi} d\theta e^{i \theta} P(\theta)  \right|.
\label{eq:sharpness}
\end{equation}
For a $\delta$-function distribution in $\theta$, the variance is 0, while for a flat distribution $V[P(\theta)] \rightarrow \infty$. One can show (see e.g.  \cite{BWB01}) that for small phase variance $V[P(\theta)] \approx \Delta^2(\theta)$ where $\Delta^2(\theta)=\langle (\theta-\langle \theta\rangle)^2\rangle$ is the usual variance.\\

Assume that we execute the circuit in Fig.~\ref{fig:ape} for $M$ rounds so that one gets an $M$-bit estimate $\tilde{\theta}$ of the phase $\theta$ and let $x[m]$ be the $m$-bit string of 0 or 1 measurement outcomes after $m$ rounds. The adapative phases will be set to the values $\varphi_1, \ldots \varphi_m$, $m=1,\ldots,M$ and the question is how these will be chosen to maximize the sharpness and thus minimize the Holevo phase variance. We assume no priori knowledge on the phase $\theta$, i.e. the initial probability distribution over $\theta$, $P(\theta)$, is assumed to be a flat distribution. This is reasonable as the squeezed (in $q$) vacuum state to which the $S_p$ measurement is applied is a superposition over eigenstates with rather uniformly distributed phases $\theta$ of $S_p$. 
If we assume no prior knowledge about $\theta$, one may as well choose the first phase $\varphi_1=0$ which is what we do.

It can then be argued, see the self-contained analysis in Appendix \ref{sec:adap_phase}, that one should choose the next phases $\varphi_2, \ldots,\varphi_M$ as follows, depending on the previous measurement outcomes:
\begin{equation}
\varphi_m= \argmax_{\varphi} \sum_{x_m=0,1}\left|\int d \theta e^{i\theta}P_{\varphi}(x[m]|\theta) \right|, 
\label{eq:feedback}
\end{equation}
where the probability $P_{\varphi}(x[m]|\theta)$ is the probability of obtaining measurement outcomes $x_1, \ldots, x_m$ given an eigenstate $\ket{\psi_{\theta}}$. This probability has a simple expression as the measurement results of each round are independent, i.e. 
\begin{equation}
P_{\varphi=\varphi_m}(x[m]|\theta)=\prod_{i=1}^m P_{\varphi_i}(x_i|\theta)=\prod_{i=1}^m \cos^2\left(\frac{\theta+\varphi_i}{2}+x_i \frac{\pi}{2}\right).
\label{eq:prod}
\end{equation}
For a (small) number of measurements, say, $M=1, \ldots, 20$, one can simply solve this expression for the optimal values for $\varphi_2,\varphi_3, \ldots$ numerically given all previous possible measurement outcomes and store these optimal values as a look-up Table which is what we have done.

Given that one has obtained a final measurement record $x[M]$, how does one choose an estimate for the phase $\tilde{\theta}$? In \cite{BWB01} it is argued that the estimated value $\tilde{\theta}$ should be chosen as 
\begin{equation}
\tilde{\theta}=\arg \int_{-\pi}^{\pi} d\theta e^{i\theta} P(\theta|x[M])  =\arg \int_{-\pi}^{\pi} d\theta e^{i\theta} P(x[M]|\theta). 
\label{eq:theta_est}
\end{equation}
Here we have used that $P(\theta|x[M])=P(x[M]|\theta) P(\theta)/P(x[M])$. Note that $P(\theta)$ is a flat distribution and $P(x[M])=2^{-M}$ which do not influence the $\arg$ function.  The probability $P(x[M]|\theta)$, which implicitly depends on the feedback phases $\varphi_1, \ldots,\varphi_m$ was given in Eq. (\ref{eq:prod}). If all feedback phases are set to 0, then $\tilde{\theta}$ will be estimated as 0 or $\pi$ as the function of which we take the argument is real: this will however not occur since $\varphi_2 \neq 0$ after the first bit has been generated. If we do not use any feedback, the value of this estimate $\tilde{\theta}$ coincides with the estimate in Eq.~(\ref{eq:theta_est_sim}).

In Fig.~\ref{fig:deviation} we plot the probability for obtaining a phase error $\delta \theta=|\tilde{\theta}-\theta|$ given a fixed number of rounds $M$, averaged over random input phases $\theta$ and runs through the protocol for both the non-adapative phase estimation protocol and the adaptive phase estimation protocol.
One important difference between the two schemes is that in the non-adaptive scheme with $\varphi=0$ and $\varphi=\pi/2$ one has a total of $(\frac{M}{2}+1)^2$ possible outcomes, as the order of the outcomes for a fixed value of $\varphi$ does not affect the measurement operator which is applied.
This is clearly not true for adaptive phase estimation. Setting $\varphi_1=0$, we have two possible values for $\varphi_2=\varphi_2^{x_1}, x_1=0,1$. Then, there are four possible values for $\varphi_3^{x_1 x_2}$ etc. as the optimization in Eq.~({\ref{eq:feedback}) depends on all previous outcomes. The final estimate for $\theta$ in Eq.~(\ref{eq:theta_est}) will depend on all these phases $\varphi_i^{x_1\ldots x_{i-1}}$ and $\tilde{\theta}$ can thus take on $2^M$ possible values. These arguments suggest that the adaptive protocol can give a more accurate estimate of $\theta$. Fig.~\ref{fig:deviation} shows indeed how adapative phase estimation (APE) outperforms such simple phase estimation by repetition (PE) for a small number of rounds $M$. This difference will become more pronounced for larger $M$, see e.g. \cite{BWB01}, but the improvement is relatively small here.

\begin{figure}[htb]
    \centering
    \includegraphics[width=\hsize]{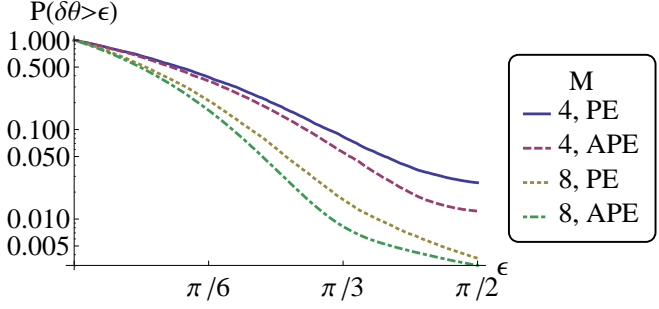} 
    \caption{(Color online) Total probability $P(\delta \theta> \epsilon)$ with $\delta \theta=|\tilde{\theta}-\theta|$ versus $\epsilon$ for $M=4,8$ (averaged over $\theta \in (-\pi,\pi]$). APE is the adaptive phase estimation protocol described in this section. PE is the non-adapative protocol where one sets the feedback phase $\varphi=0$ for $M/2$ rounds and $\varphi=\pi/2$ for $M/2$ rounds. For $\delta \theta < \pi/3$, one is below the shift error threshold.}
    \label{fig:deviation}
\end{figure}

In order to understand better how well these phase estimation methods project the input state onto an approximate eigenstate, we explicitly numerically generate the states that are created through these protocols in Section \ref{sec:num}.  

\begin{figure}[htb]
\subfigure[][]{
\includegraphics[width = .45\hsize]{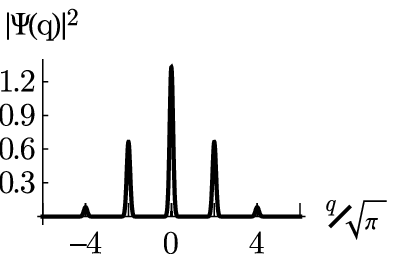}
} \subfigure[][]{
\includegraphics[width = .45\hsize]{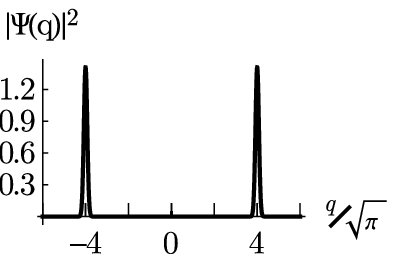}
}
\caption{%DW RESUBMIT CHANGE: removed color, "color online".
Absolute values of two examples of wave functions of code states which can be generated by non-adaptive phase estimation for $M=4$ and $\Delta=0.2$. After the eigenvalue measurement, the states are shifted onto an approximate code state using the phase estimate. The left state (a) is obtained if the first and the second measurement of both $U$ and $iU$ yield the same result, the right state (b) is obtained if both results differ.}
\label{fig:wavefunctions}
\end{figure}

\subsection{Preparing the Input State: Two Methods}

One has two alternatives for preparing a code state such as $\ket{0}$ (or $\ket{+}$) approximately. As we have argued, one can start the protocol in a sufficiently squeezed vacuum state $\ket{\rm sq. vac}$ and approximately measure $S_p$ (or squeeze in the other quadrature and measure $S_q$ to prepare $\ket{+}$). Since the squeezed vacuum state is an approximate eigenstate of $Z$ one produces an approximate $\ket{0}$ state. 
How many dB of squeezing does one need in order to get a good eigenstate of $Z$? Let the squeezed vacuum state be $\ket{{\rm sq. vac.}}=\int \;dq \;\alpha_q \ket{q}$, Eq.~(\ref{eq:sqvac}), with $\Delta=e^{-r}$ and $\alpha_q=\frac{1}{\sqrt[4]{\pi}}e^{r/2} e^{-q^2 e^{2r}/2}$. 
The probability for shift errors above the shift error threshold is given by $P_{\rm error,q}^{\sqrt{\pi}/6} \leq \int_{-\sqrt{\pi}/6}^{\sqrt{\pi}/6} dq \; |\alpha_q|^2={\rm Erf}(e^r \sqrt{\pi}/6)$. One sees in Fig.~\ref{fig:squeeze} that the approach to $100\%$ success probability happens around 8 dB of squeezing. 
%BMTcor
Squeezing of itinerant microwave fields, instead of a confined cavity field, can be achieved using a Josephson parametric amplifier (JPA). For example, in \cite{mallet+:squeeze} one obtains $\Delta^2=(\Delta q_{\rm sq})^2/(\Delta q_{\rm vac})^2 \approx 12\%$ corresponding to $\Delta\approx 0.35$ and squeezing around 4.1 dB. More recent work has achieved squeezing of at least 10 dB \cite{cast2008}.

A more recent proposal was considered in \cite{EG:squeezing}: in this paper it is analyzed how a squeezed microwave drive can be used to produce a squeezed vacuum state as the stationary state of the cavity field (under dissipative photon loss dynamics). 

% Delta(x)^2 is 12% of vacuum variance
% G=cosh^2(r) or G=10 log ( 2 Delta X^2) =10 log (exp(2r)) convention USED ?? what is the gain...
% 10 log_10 G=
% $\Delta x=\frac{1}{\sqrt{2}}$. 
% Delta=exp(-r)

\begin{figure}[htb]
    \centering
    \subfigure[][]{
    \includegraphics[width=.45\hsize]{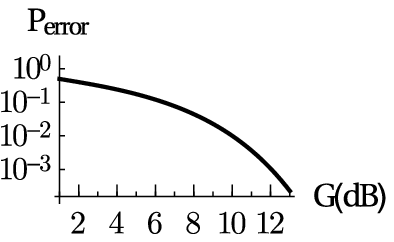}
        }
    \subfigure[][]{
    \includegraphics[width=.45\hsize]{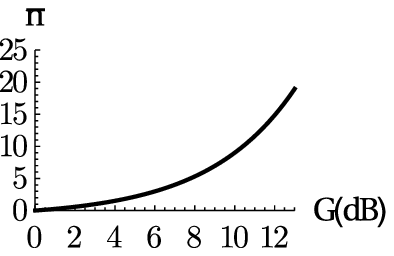} 
    }
\caption{%DW RESUBMIT CHANGE: removed color, "color online".
(a): Upper bound on the effective error probability $P_{\rm error}=P_{\rm error,q}^{\sqrt{\pi}/6}$ for a squeezed vacuum state in the \emph{squeezed} quadrature. (b): The average number of photons in the squeezed vacuum state depending on the gain $G$ in dB. }
\label{fig:squeeze}
\end{figure}

An alternative is to start the entire protocol in the vacuum state of the cavity mode and execute the phase estimation protocol both for $Z$ as well as $S_p$. The approximate measurement of $Z$ is thus also an effective means of producing a squeezed state from the vacuum state in the cavity mode. The phase uncertainty $\delta \theta$ in the estimate of the eigenvalue of $Z$ needs to be at most $\pi/6$ in order to be below the shift error threshold (instead of $\pi/3$), {\em twice} as small as compared to the $S_p$ measurement. 
 This suggests that if the phase of $S_p$ is measured in $M$ rounds, then the phase of $Z$ should be measured in $4M$ rounds in order to give an overall similar shift error contribution. As $Z=D(i \sqrt{\pi/2})$ as compared to $S_p=D(\sqrt{2\pi})$ the average number of photons added per $Z$-round is $1/4$ of a $S_p$-round, hence the total contribution to the number of photons would be about the same. These arguments show that it may be most advantagous to start with a squeezed vacuum state instead of selecting the code state by phase estimation of $Z$  as the $Z$-measurement would be rather lengthly (at least $32$ rounds for $M  \geq 8$) during which the code state is also decohering.
At the same time, the $Z$-phase estimation protocol by itself may be an interesting novel way to prepare a squeezed state.

\subsection{Numerical Analysis of Two Phase Estimation Schemes for $S_p$}
\label{sec:num}

We apply the nonadaptive and adapative phase estimation scheme for $S_p$ with $M$ rounds to a squeezed vacuum state with parameter $\Delta$ in order to create an approximate code state $\ket{0}_{\rm approx}$. The state obtained through the sequence of measurements is a superposition of displaced squeezed vacuum states, each with a phase which depends on the feedback phase and the measurement outcome $x[M]$:
\begin{equation}
\ket{\Psi(x[M])} = \frac{1}{\sqrt{N}}\sum_{j=0}^{M} c_j(x[M]) D(\sqrt{2\pi}(j-M/2))\ket{\rm sq. vac},
\label{eq:state_x}
\end{equation}
with normalization $N \approx \sum_{j=0}^{M} | c_j(x[M]) |^2$ (the approximation in the normalization comes from $\bra{\rm sq. vac} D(\sqrt{2\pi}(t-t')) \ket{\rm sq. vac}\approx 0$ for $|t-t'| \geq 1$). Here the coefficients $c_j(x[M]) =\sum_{ \{\mathcal{S}_j\} } \prod_{k \in \mathcal{S}_j} e^{i(\varphi_k+x_k \pi)}$ where the subsets $\mathcal{S}_j$ are subsets of $j=|{\cal S}_j|$ indices chosen from $1,\ldots,M$ (without repetition). There are ${M \choose j}$ such subsets ${\cal S}_j$. Also, $\varphi_k$ is the feedback phase for round $k$ and $x_k$ is the outcome bit of round $k$.
From the phase estimation one infers a value $\tilde{\theta}\in (-\pi,\pi]$, upon which one corrects $\ket{\Psi(x[M])}\rightarrow \ket{0(x[M])}_{\rm approx}=e^{i v q} \ket{\Psi(x[M]]}$ with $v=\frac{\tilde{\theta}}{2  \sqrt{\pi}}$. Note that the additional displacement to center the code state around the vacuum does not affect the eigenvalue of $S_p$ as it commutes with $S_p$.  

A few examples of the wave functions of states that one can obtain are shown in Fig.~\ref{fig:wavefunctions}. These examples show that one is not always guaranteed to get a good wavefunction and the effective probability of error can vary depending on the measurement outcomes. Because the relation between the effective error rate and the measurement result is known, it is possible to use heralding or effectively use some post-selection. If we know (by simulation) that the obtained code state is bad, one would repeat the protocol. 

In Fig.~\ref{fig:ape_hist} at the end of the paper we plot the total probability $P$ to prepare a code state with effective error rate $P_{\text{error,p}}^{\sqrt{\pi}/6}$ for the (adaptive) phase estimation. As the effective error rate due to finite squeezing and the effective error rate due to the phase estimation scheme are unrelated (one corresponds to shifts in $q$, while the other corresponds to shifts in $p$),
%DW RESUBMIT CHANGE:
%REMOVED: the simulation is done with an infinitely squeezed state as initial state.
%REASON: Not true with the updated simulation, the new simulation works on finite squeezing, but separates the two contributions in the analysis. I checked explicitely in the simulation that the figure is indeed independent of the finite squeezing. 
only the part $P_{\text{error,p}}^{\sqrt{\pi}/6}$ due to the phase estimation scheme is shown. 
%NEW TEXT ABOVE 
  The effective error rate including both effects can be computed using $P_{\text{error}}^{\sqrt{\pi}/6} = P_{\text{error,p}}^{\sqrt{\pi}/6}+P_{\text{error,q}}^{\sqrt{\pi}/6}-P_{\text{error,p}}^{\sqrt{\pi}/6}P_{\text{error,q}}^{\sqrt{\pi}/6}$. The error rate $P_{\text{error, q}}^{\sqrt{\pi}/6}$ is shown in Fig. \ref{fig:squeeze}.

Both protocols show a moderate probability to obtain code states with low error rates.
%DW RESUBMIT CHANGE: it is 94% for APE and 68% for PE. "Much more" is still ok, I think. I changed the number.
What is striking is that the adaptive version of the protocol is much more reliable than the nonadaptive version and produces code states with an effective error rate below $1\%$ in up to $94\%$ of the cases.

It is noteworthy that in most cases, both protocols require some heralding/post-selection, as there is a finite probability to prepare a code state with a large effective error rate.
It should also be clear that using phase estimation in this way is much superior to simply post-selecting on obtaining the outcome $x_1=x_2=\ldots=x_M=0$ which occurs with an exponentially small probability in $M$. If phase estimation is used for quantum error correction, then post-selection is not an option, as we do not wish to throw away the encoded state.

\subsubsection{Number of Photons}
\label{sec:photon}

We also consider the mean number of photons and fluctuations thereof for the adaptive and nonadaptive phase estimation of $S_p$, applied to a squeezed vacuum state with parameter $\Delta$. The expected number of photons in $\ket{\Psi(x[M])}$ (which is only slightly different from the number of photons in $\ket{0(x[M])}$) approximately equals
\begin{equation}
\overline{n} (x[M]) \approx \sinh^2(\ln(\Delta)) + \frac{1}{N  }\sum_{j=0}^{M} | (j-M/2) c_j(x[M])  \sqrt{2\pi}|^2,
\label{eq:nav_x}
\end{equation}
where we again used that $\bra{\rm sq. vac} D(\sqrt{2\pi}(t-t')) \ket{\rm sq. vac}\approx 0$ for $|t-t'| \geq 1$ and the normalization $N$ has been given above. In Fig.~\ref{fig:squeeze} we plot the contribution from squeezing on the mean photon number.
%DW RESUBMIT CHANGE
%REASON: fig:photon now shows the total contribution, as the photon number is n(x,\Delta) and cannot be easily separated
 In Fig.~\ref{fig:photon}  we plot the {\em total photon number} after the measurement of $S_p$. There, we see $\langle \overline{n}\rangle_{\theta}=\frac{1}{2\pi}\int d\theta P(x[M]|\theta) \,\overline{n}(x[M])$ with $\overline{n}(x[M])$ as in Eq.~({\ref{eq:nav_x}) as a function of $M$. We also plot the standard-deviation 
$\sqrt{\langle (\overline{n}-\langle \overline{n})^2\rangle_{\theta}\rangle_{\theta}}$.  Note that this is not the standard-deviation
$\sigma(n)$ of the state itself which, as we discussed in Section \ref{sec:intro}, scales as $\overline{n}$.
%DW RESUBMIT CHANGE: two new sentences below
 As the contributions from squeezing and the phase estimation cannot be separated for finite squeezing, we show the photon numbers for $\Delta = 0.2$. Note that both mean value and deviation depend on $\Delta$.
 %DW RESUBMIT CHANGE: modified formula, changed to below 50
One can observe that $\overline{n} \approx M \pi/2 +\overline{n}_{\text{squeeze}}$: this is what we expect as in each round we symmetrically displace the input state further  out to the left and right by an amount $\alpha=\sqrt{\pi/2}$, increasing the expected photon number thus by $\pi/2$.The total expected number of photons in the cavity for $M=8$ and $\Delta=0.2$ ($8.3$ dB) is below 25 such that $\overline{n}\pm \sigma(n)$ is below 50.

\begin{figure}[htb]
    \centering
    \subfigure[][]{
    \includegraphics[width = .4\hsize]{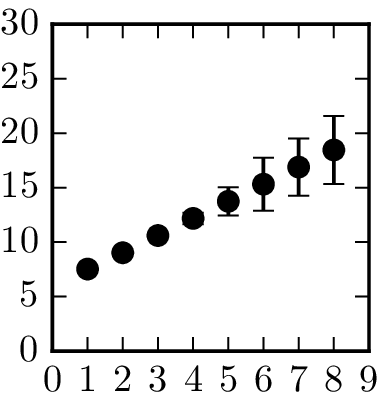}
    }
    \subfigure[][]{
    \includegraphics[width = .4\hsize]{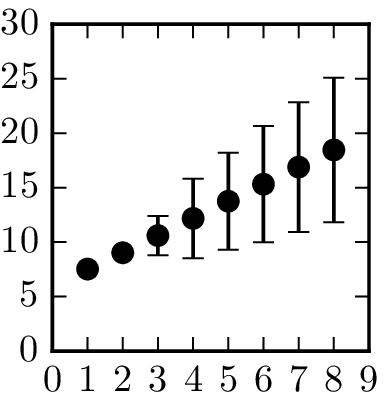}
    }
    \caption{%DW RESUBMIT CHANGE: removed color, "color online".
    %CHANGE: Figure now shows total photon number, as \bar{n} can only be separated in squeezing and phase estimation for infinite squeezing. This changes both mean and variance.
    The average number of photons $\langle \overline{n}\rangle_{\theta}$ after (a) adaptive phase estimation of $S_p$ and (b) non-adaptive phase estimation of $S_p$ using $M$ ancilla qubits (rounds) with parameter $\Delta = 0.2$. The fluctuations $\sqrt{\langle (\overline{n}-\langle \overline{n})^2\rangle_{\theta}\rangle_{\theta}}$ around the average number of photons reflect the different measurement outcomes in the protocol. It is noteworthy that these fluctuations are much smaller for the APE than for the PE protocol.
}
    \label{fig:photon}
\end{figure}

\subsection{Usage of Adaptive Phase Estimation in Quantum Error Correction}
Adaptive phase estimation of $S_p$ and $S_q$ can in principle be used for quantum error correction. One cycle of quantum error correction will then consist of phase estimation of $S_p$ followed by phase estimation of $S_q$, each taking a certain number of $M$ rounds.
Each time one starts measuring $S_p$ one has prior information about its phase which should be used in choosing the next feedback phases. If the entire QEC protocol is noiseless, then the choice for the feedback phase of the first round of the $S_p$ measurement is simply estimated using the previous feedback phases and outcomes of the previous measurement of $S_p$, as determined by Eq.~(\ref{eq:feedback}). It is clear that for such a noiseless protocol, the number of photons thus keeps increasing round by round while the state gets closer and closer to a true eigenstate of $S_p$ (and $S_q$).
In the implementation, see Section \ref{sec:prop}, the number of photons is limited and at high number of photons the interactions that are invoked to execute the protocol are no longer accurate. In addition, photon loss is occuring continuously from the cavity during the protocol.
This means that for quantum error correction the choice of feedback phases should ideally take into account an error model of the dynamics during the $S_q$ measurement and how this dynamics changes the current estimate phase for $S_p$. We leave such more complete analysis to future work.

\section{Proposal for Realization in Dispersive Qubit-Cavity Set-Up}
\label{sec:prop}

We consider the following physical set-up. A superconducting transmon qubit is capacitively coupled to a 2D or 3D microwave high-Q storage cavity, as well as a low-Q read-out cavity. The storage cavity will be used to prepare a code state and the read-out cavity will be used to measure the state of the qubit. This is the same set-up as the cat state code experiments in e.g. \cite{vlastakis+:cat100} and we will use values for physical parameters which are similar as in this setup, see Table \ref{table:parameters}.
We assume that a particular cavity mode $a$ with resonance frequency $\omega_r$ couples most strongly to the transmon qubit and neglect the interactions of the qubit with other cavity modes, as well as the coupling to all higher-energy levels beyond the states $\ket{0}$ and $\ket{1}$.

We assume that the interaction between qubit and storage cavity mode is approximately described by a simple Jaynes-Cummings Hamiltonian $H_{JC}=-\frac{\omega_q^{\rm bare}}{2} Z+\omega_r a^{\dagger}a+g (\sigma_- a^{\dagger}+\sigma_+ a)$. In the dispersive regime when $\frac{g}{\Delta} \ll 1$ ($\Delta=\omega_q^{\rm bare}-\omega_r$ is the detuning), one can make a perturbative expansion in $\frac{g}{\Delta}$ and derive an effective Hamiltonian which equals
\begin{eqnarray}
H_{\rm eff}=(\omega_r-\chi Z)a^\dagger a-\frac{1}{2} \omega_q Z+O\left(\frac{g^4}{\Delta^3}\right), 
\label{eq:heff}
\end{eqnarray}
with dispersive shift $\chi \approx \frac{g^2}{\Delta}+O(g^4/\Delta^3)$. Here $\omega_q=\omega_q^{\rm bare}+\chi$ where $\omega_q^{\rm bare}$ is the bare qubit frequency when the qubit is uncoupled to the cavity.   
The perturbative expansion is warranted when $\overline{n} < n_{crit}=\Delta^2/(4g^2)$ where $\overline{n}$ is the mean number of photons in the cavity.
The effective Hamiltonian shows that the resonant frequency of the cavity is shifted depending on the state $\ket{0}$ ($+$) or $\ket{1}$ ($-$) of the qubit, i.e. its frequency $\omega_r \rightarrow \omega_r^{\pm}=\omega_r \mp \chi$. In this approximation we neglect a nonlinear term in the effective Hamiltonian of the form $\frac{5 g^4}{3\Delta^3} Z (a^{\dagger} a)^2 \equiv \chi' Z (a^{\dagger} a)^2$, a nonlinear dispersive shift.

In a more systematic approach the qubit-cavity Hamiltonian can be obtained through first determining the normal modes of the coupled $LC$ circuits, after which the nonlinearity due to the Josephson junction is treated as a perturbation \cite{nigg+:bb}. This gives rise to an effective Hamiltonian of the form 
\begin{equation}
H_{\rm eff}=\omega_r a^{\dagger} a+\tilde{\omega}_q b^{\dagger} b-\frac{\chi_{rr}}{2} (a^{\dagger} a)^2-\frac{\chi_{qq}}{2}(b^{\dagger} b)^2-\chi_{qr} a^{\dagger}a b^{\dagger} b,
\label{eq:heff2}
\end{equation}
with $\chi_{rr}\equiv 2K=\frac{\chi_{qr}^2}{4 \chi_{qq}}$. Thus both qubit and cavity are represented as coupled nonlinear oscillators such that the anharmonicity for the qubit $\chi_{qq}$ is relatively large. When one restricts the $b$-oscillator to the lowest two levels, one can identify $\omega_q \approx \tilde{\omega}_q-\chi_{qq}/2$ and $\chi_{qr} \approx 2\chi$. Higher-order terms in Eq.~(\ref{eq:heff2}) of the form $(a^{\dagger} a)^2 b^{\dagger}b$ would describe the nonlinear dispersive shift. Note that in the model in Eq.~(\ref{eq:heff}) the cavity self-Kerr nonlinearity is not present while its non-zero value has been determined in the experiments.

The coupling of the qubit to the read-out cavity is described by a similar effective Hamiltonian as in Eq.~(\ref{eq:heff}) with a weaker dispersive coupling. \\

 In Table \ref{table:parameters} we give the ranges of the relevant physical parameters and we choose several specific values in these ranges to demonstrate how well the protocol can be executed. We will take $\chi/2\pi=2.5$ MHz and let the detuning be $\Delta/2\pi=1$ GHz in which case $n_{\rm crit} \approx \frac{\Delta}{4 \chi}=100$. This upper limit on the number of photons is sufficient for creating good code states: for an 8-round protocol $M=8$, $\overline{n} \approx 25$, see Fig.~\ref{fig:photon}, so that $\overline{n} \pm \sigma(n) \lessapprox 50$ (with $\sigma(n) \sim\overline{n}$), well below this limit. For this choice of parameters $g/2\pi \approx 50$ MHz. The strength of the neglected nonlinear term in Eq.~(\ref{eq:heff}) then equals $\chi'/(2\pi)\approx \frac{5 g^4}{6 \pi \Delta^3}=10$ kHz. In the experiment described in \cite{vlastakis+:cat100} in which cat states are created with about 55 photons, a value for $\chi'/(2\pi) \approx 4.2$ kHz is estimated for a detuning $\Delta/2\pi=1.18$ GHz and $\chi/2\pi=2.4$ MHz which is considerably lower. The strength of the self-Kerr nonlinearity $K  (a^{\dagger} a)^2$ in this experiment is estimated as $K/(2 \pi) \approx 3.61$ kHz. The authors estimate $n_{\rm crit} \leq 300$ which is a bit more optimistic than the estimate above.

% comment on low loss and needed coupling to outside world, so more microwave power

\begin{table*}[htb]
\centering
\begin{tabular}{ll}
Transmon qubit $\frac{\omega_q}{2\pi}$ and bare cavity frequency $\frac{\omega_r}{2\pi}$ & $3-11$ GHz\\
Qubit $T_1/T_2$ time & $10-100\,\mu$sec   \\
3D (storage) cavity lifetime $T_{\rm cav}$ & $55\,\mu$sec \cite{sun+:cat}, $1$ msec ($Q > 10^7$) \cite{RCS13} \\
2D CPW (storage) cavity lifetime $T_{\rm cav}$ &  $200\,\mu$sec  ($Q > 10^6$) \cite{BDD15} \\
Controlled-displacement pulse time $T_{pulse}$ & $25-100$ nsec \cite{leghtas+:map} \\
Dispersive shift  $\chi/2\pi$ & $1-20$ MHz \\
Qubit measurement time $t_{\rm meas}$ &  $200-300$ nsec \cite{hatridge:science, sun+:cat} \\
Single qubit gate & $5-10$ nsec
 \end{tabular}
 \caption{Ranges of some relevant parameters. The quality factor values (and lifetimes) represent the internal losses inside the cavity while in our protocol the total quality factor of the cavity $Q_{\rm tot}$, including its intended but flexible coupling to the outside world, is the relevant parameter.}
 \label{table:parameters}
\end{table*}

% lifetime=2 pi/kappa =2 pi Q /omega

One round of phase estimation will last a total time $T_{\rm round}$. During this time a quantum circuit depicted in Fig.~\ref{fig:rpe} is to be executed.
The protocol of a round consists of a short (O(10)ns) interval $T_{\rm prep}$ in which qubit (and storage cavity) are prepared. For the qubit this means it is put in the state $\frac{1}{\sqrt{2}}(\ket{0}+\ket{1})$ (by a Hadamard of $R_x(\pi/2)$ gate). After $T_{\rm prep}$, qubit and storage cavity should be coupled by a controlled-displacement transformation or a $D(Z\sqrt{\pi}/2)$ gate (in phase estimation for $S_p$).
For this gate there are two options, as has been discussed in \cite{leghtas+:map,vlastakis+:cat100}.\\

In some physical set-ups the dispersive cavity-qubit coupling (both storage and read-out cavity) is not tunable and is thus always `on'. Such a setup is non-ideal in various ways. When the dispersive coupling is always on, it means that one should prepare the code states in a rotating frame (not the lab frame) which depends on the qubit state. For example, one chooses the frame in which the qubit is in the state $\ket{0}$ as the frame in which the cavity state should be unchanging, stationary (and we also look at the qubit in its rotating frame as gates on the qubit are done relative to that). In this frame the effective Hamiltonian in Eq.~(\ref{eq:heff}) equals $\tilde{H}_{\rm eff}=2 \chi \ket{1}\bra{1} a^{\dagger}a$. It is important that during the qubit measurement, which takes up a considerable amount of time, this effective Hamiltonian induces no further rotations on the partially-prepared code state (or if it induces rotations, one should know what they are). This means that during qubit measurement the qubit has to quickly be reset to $\ket{0}$, in order to induce no further rotational dynamics on the cavity state. 

Another disadvantage of using a non-tunable $\chi$ is that the accuracy of single qubit rotations depends on the number of photons in the cavity. The qubit frequency given a storage cavity with $n$ photons is, by Eq.~(\ref{eq:heff}), given by $\omega_q+2 \chi n$. A microvave pulse which should rotate the qubit independent of the number of photons in the cavity should thus qualitatively take at least time $T_{\rm pulse}=2/W$ \cite{BDT:circuitQED} with frequency width $W \geq 2\chi \sigma(n)$. This assumes that one sets the center frequency of the pulse at $\omega_q+2 \chi \overline{n}$. Here $\sigma(n)$ is the standard deviation in a code state, given in Eq.~(\ref{eq:dev}), which scales with $\overline{n}$. In \cite{vlastakis+:cat100} it was argued that the unwanted entangling of qubit and cavity due to single qubit rotations is a leading source of inaccuracies when one goes to higher photon numbers. Even though the number of photons in our proposed protocol for $M=8$ will never be larger than 50, the protocol consists of many more single qubit gates than the experiments in \cite{vlastakis+:cat100} so that these errors will accumulate.
With a non-tunable $\chi$, the dispersive coupling to the read-out cavity is of course continuously on during the entire round while in the Figures it is suggested that measurement only occurs at the end of the protocol. It is understood that the coupling between transmon qubit and read-out cavity mode is smaller than the coupling between transmon qubit and storage cavity, for example in \cite{sun+:cat} the dispersive coupling $\chi/2\pi$ to the readout cavity was estimated as $0.930$ MHz. At the same time, this coupling needs to sufficiently strong to provide a relatively short measurement time for the qubit as $T_{\rm round}$ will be largely determined by the length of the controlled-displacement transformation and the qubit measurement time. \\
A third disadvantage of a non-tunable $\chi$ is the the cavity Kerr nonlinearity which is present in Eq.~(\ref{eq:heff2}) due to the linear coupling between the LC oscillators: the `cavity mode' is in fact a `dressed' cavity mode which sees the Josephson nonlinearity. If $\chi$ is turned to a small value, then this Kerr nonlinearity will be correspondingly small. \\

For these reasons, we imagine that the dispersive coupling $\chi$ is tunable and can be turned `on and off' during the execution of a round. Thus during each round, the dispersive storage cavity-qubit coupling $\chi(t)$ is `off' (i.e. to a low value) during qubit preparation and single qubit rotation (lasting time $T_{\rm prep}$) and possibly unconditional cavity displacements (of the form $D(-\sqrt{\pi/2})$). Then $\chi$ is rapidly turned on after $T_{\rm prep}$ and stays on for a time-interval $T$ during which the controlled-displacement gate in Fig.~\ref{fig:rpe} acts. The coupling $\chi$ is again rapidly turned off during the last interval $T_{\rm post}$ during which the qubit state undergoes further single qubit rotations and is being measured.  One can similarly imagine that only during the measurement time the coupling to the read-out cavity is turned on.
The turning off and on of $\chi$ could be achieved in two possible ways. One can increase the cavity-qubit detuning $\Delta$ (and hence reduce $\chi$) by altering the qubit resonant frequency $\omega_q^{\rm bare}$ using a flux-tunable transmon qubit; such switching can take place in $O(1)$ nsec. 
An alternative is to have a variable qubit-cavity capacitive coupling \cite{srinivasan+:tunable, hoffman+:tunable} which is turned on and off in a $O(1)$ nsec time window.\\

Let us now discuss the realization of the controlled-displacement gate. Note that if $\chi$ is in principle turned `off'  unless the cavity state needs to be manipulated, it means that we are preparing the code states in the rotating frame of the cavity at frequency $\omega_r$ (while single qubit gates are performed in the frame rotating at $\omega_q^{\rm bare}$). When the coupling is turned on, the effective Hamiltonian in these rotating frames is then $\tilde{H}_{\rm eff}=-\chi Z a^{\dagger} a-\frac{1}{2} \chi Z$ (neglecting nonlinearities). The additional $Z$-rotation in this Hamiltonian leads to a phase accumulation on the prepared qubit state $\frac{1}{\sqrt{2}}(\ket{0}+\ket{1})$ which should be taken into account when considering what single qubit $Z$-rotation of the form ${\rm diag}(1,\exp(i \varphi))$ is done during $T_{\rm post}$.\\

In principle, one can enact a controlled-displacement gate by supplementing this dynamics by driving the cavity with a microwave pulse. During the time interval $T$, a microwave drive of duration $T_{\rm pulse} \approx T$ is applied. The drive tone of this pulse $\omega_d=\omega_r+\chi$ is resonant with the cavity mode when the qubit is in the state $\ket{1}$ but off-resonant when the qubit is in state $\ket{0}$. Hence, one expects a large cavity displacement in the resonant case and a small neglible displacement when the drive-tone is off-resonant, thus enacting a controlled-displacement gate. Again, at an intuitive level, the frequency width $W$ of the pulse should be sufficiently narrow as compared to $2\chi$, which is the difference in resonance frequencies of the oscillator given the qubit state, so that the pulse has few photons at frequency $\omega_r-\chi$. If we center the pulse at $\omega_r+\chi$, then for a Gaussian pulse with $T_{\rm pulse}=2/W$, one should at least require that $2 \chi > \frac{W}{2}=1/T_{\rm pulse}$ or $\chi T_{\rm pulse} > 1/2$. This heuristic argument assumes that the cavity decay rate $\kappa$ which sets the width of the resonance obeys $\kappa \ll W$ which is warranted for high-Q storage cavities. 

During this pulse, the qubit-cavity coupling $-\chi Z a^{\dagger} a$ also induces a qubit-state dependent phase space rotation on the cavity state, i.e. of the form $\exp(i \chi Z a^{\dagger}a t) $, which is in fact {\em undesired}, see Fig.~\ref{fig:phasespace_sketch} in Appendix \ref{sec:direct}. It is thus important to choose the interaction-time $T$ such that this difference in rotation angle is $2\pi$ which implies that the newly created superposition of displaced squeezed states stays on a line. This warrants a precise choice for $T$ namely $\chi T=\pi$. In the analysis in Appendix \ref{sec:direct} we model the drive and the corresponding displacements and rotations assuming a simple square pulse (with zero rise time). The conclusion of this analysis is much less negative than what has been stated in \cite{leghtas+:map} where it is written that a direct controlled-displacement gate requires a time $T_{\rm pulse} \approx 30/\chi$ which would be quite long.

The alternative is to do a controlled-displacement gate through a sequence of two controlled-rotations interspersed with an unconditional displacement \cite{vanloock:hybrid}, explicitly shown in Fig. \ref{fig:cont_disp}.

\begin{figure*}[htb]
      \centering
      \begin{tabular}{ccccc}
%	\Qcircuit @C=1em @R=.7em {
%	&&&&\lstick{\text{cavity}} & \qw &  \multigate{1}{R(-Z\pi/2)} & \gate{D(-i\alpha/2)} &\qw & \multigate{1}{R(-Z\pi/2)} & \qw & \qw  \\
%	&&&&\lstick{\text{qubit}}  & \qw & \ghost{R(-Z\pi/2)} & \gate{X}  & \qw& \ghost{R(-Z\pi/2)}  & \qw &  \gate{X}
%	}
	\includegraphics{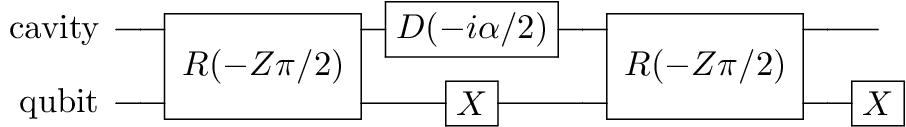}&
    $=$&
%	\Qcircuit @C=1em @R=.7em {
%	& & \multigate{1}{D(-Z\alpha/2)} & \qw \\
%	& &  \ghost{D(-Z\alpha/2)}& \qw & \push{\vphantom{X}}
% 	}
	\includegraphics{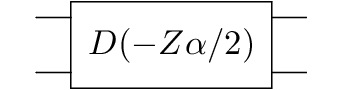}&
    $=$ &
%	\Qcircuit @C=1em @R=.7em {
%	& \gate{D(-\alpha/2)}   & \gate{D(\alpha)} & \qw \\
%	& \qw &  \ctrl{-1}& \qw & \push{\vphantom{X}}
% 	}
	\includegraphics{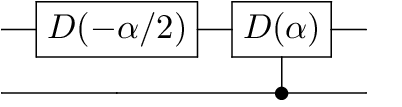}
	\end{tabular}
    \caption{Realization of a controlled-displacement gate via two controlled-rotations for one round of phase estimation of $S_p$ with $\alpha=\sqrt{2\pi}$ or $S_q$ with $\alpha=-i\sqrt{2\pi}$, see Fig.~\ref{fig:rpe}. Here $R(-Z \pi/2)=\exp(i a^{\dagger}a Z \pi/2)$.}
    \label{fig:cont_disp}
\end{figure*}

Let $R(\theta)=\exp(-i\theta a^{\dagger} a)$ be a rotation in phase space. The conditional rotation $R(-Z \pi/2)=\exp(i a^{\dagger} a Z \pi /2)$ is obtained by $\tilde{H}_{\rm eff}$, i.e. coupling the transmon qubit for time $t$ with $t \chi=\pi/2$ to the cavity mode. The inverse rotation equals $R(Z \pi/2)= X R(-Z \pi/2) X$ where $X$ is a Pauli $X$ ($\pi$-rotation) on the qubit. The advantage of this way of doing a controlled-displacement is that its implementation is always fast, as it requires $O(10)$ nsec. unconditional displacements and single qubit-rotations as well as a total qubit-cavity interaction time of $ T=\pi/\chi$ during which the cavity state is rotated depending on the qubit state. For our chosen $\chi$, one has $T=\pi/\chi=200$ nsec.
Note that the last $X$-gate on the qubit can be absorbed in the post-processing stage during which one thus applies a single qubit rotation which is composed of Pauli $X$ in Fig.~\ref{fig:cont_disp}, the possible feedback rotation ${\rm diag}(1,\exp(i\varphi))$, including corrections for accumulated phases, and then a Hadamard gate, shown in Fig.~\ref{fig:rpe}.
However, in order for this implementation to work, one needs to turn off the dispersive coupling $\chi$ in the middle period during which an unconditional displacement and a single qubit rotation act. Thus this scheme would require one to turn $\chi$ on (and off) twice during a round, while a direct conditional displacement would allow one to turn $\chi$ on and off once.
 \\

In either implementation of the controlled-displacement gate, $T_{\rm round}$ will be largely determined by the duration of the controlled-displacement gate and the qubit measurement time. If we assume that the qubit measurement time is approximately $300$ nsec, we have $T_{\rm round} \approx 500$ nsec. During this time qubit decoherence (see Table \ref{table:parameters}) is negligible as $T_{\rm round}/T_1(=50\mu{\rm sec})=1 \times 10^{-2}$.  
If we assume a 3D storage cavity lifetime of $1$msec (see Table \ref{table:parameters}), then during the execution of $M=8$ rounds for the full phase estimation of $S_p$, in total lasting $T_{\rm PE}=4\,\mu$sec, one has $T_{\rm PE}/T_{\rm cav}=4 \times 10^{-3}$. One cycle of quantum error correction using the measurement of $S_p$ and $S_q$ will thus last $T_{\rm QEC-cycle}=8\,\mu$sec, still considerably faster than the cavity decay time. 

For 3D cavities, no flux-tunable transmon qubits have been used so far. For a 2D co-planar waveguide cavity, to which transmon qubits have been coupled with a tunable $\chi$, the best lifetime can be $200\,\mu$sec (see Table) which would give estimates which are factor of $1/5$ worse as compared to a 3D cavity. \\

The adaptive phase estimation protocol uses feedback which implies that the phase in a single qubit rotation in the next $500$ nsec round depends on the outcome of the qubit measurement in the previous round. In current technology it is possible for a qubit measurement outcome to determine the execution of a single qubit gate $200$ nsec later, so this does not pose a problem in the implementation.

 \subsection{Shift Error and Noise During The Protocol}
\label{sec:noise}

There are two sources of errors during a round of the protocol, namely errors on the transmon qubit and direct errors on the cavity mode. We first consider errors on the transmon qubit which can propagate to the cavity mode in two ways. 

We can call a protocol strictly fault-tolerant if an error with low probability (or amplitude) on the qubit can induce a shift error $u$ or $v$ with low probability/amplitude {\em and} low strength $|u|, |v|\ll \sqrt{\pi}/2$ on the cavity mode. 

An error on the qubit during a round can alter the measurement outcome of this qubit (for example an error which flips the measurement outcomes). This will lead to an incorrect estimate for $\tilde{\theta}$. However, if the rate at which these errors take place is sufficiently low, so most rounds are error-free, then it should be clear that the error on the estimate is also small, hence the induced shift error is small,  implying fault-tolerance. One can see this as fault-tolerance which arises from the repetition of the rounds; how well this works will depend on whether the rate of qubit errors is sufficiently low.

Another type of error on the qubit can directly result in a shift error on the cavity mode.
For example, during the execution of the direct controlled-displacement or rotation, the qubit state decays to $\ket{0}$, resulting in a displacement error which can be as large as $D(\sqrt{2\pi})$. Such error shows that the phase estimation protocol is not strictly fault-tolerant, since, even though this error is very unlikely, it does lead to a large incorrectable shift error. 
Using several, say $k$, ancilla qubits per round and putting them into a cat state $\ket{00\ldots 0}+\ket{11\ldots1}$ so that each one of the $k$ qubits is only used to perform a much smaller controlled displacement $D(\sqrt{2\pi}/k)$ can mitigate this problem and make the protocol more fault-tolerant. Effectively, we are trading a low probability for a large shift error to a larger probability (as one uses more qubits) to have only small correctable shift errors. 

Given the fact that the accuracy of single qubit gates and controlled-rotations is quite high and qubit decoherence on the time-scale of a round very low, we do not anticipate that qubit errors are the dominant source of errors.\\

Let us next consider direct errors on the cavity mode which can result from photon loss, a self-Kerr nonlinearity or a nonlinear dispersive shift. In principle, for short enough time-intervals all these processes can be expanded as linear combinations of shift errors and these shift errors can be propagated through the ideal circuit in a round, or in between rounds, such that they remain shift errors which gradually add up in strength. This is true as the ideal circuit only contains (conditional) displacements and (conditional) rotations. 

Given the points made in Section \ref{sec:phys_shift}, one can consider the effect of cavity decay during preparation and the presence of nonlinearities which we expect to be the dominant source of shift errors. For cavity decay with rate $\kappa$ with $\kappa=1/T_{\rm cav}$ of a 3D cavity, one can easily meet the condition $\delta\equiv \kappa T_{\rm PE} \approx 4 \times 10^{-3}  \ll \frac{1}{n_{\rm max}}$ when $n_{\rm max}$ created during the 8-round protocol is no more than 50 photons. The arguments in Section \ref{sec:phys_shift} then imply that cavity decay modeled by some Lindblad equation induces dimensionless shift errors $u,v$ of strength $|u|,|v|\leq O(\sqrt{\kappa T_{\rm PE}})\approx 0.06$ (using only the first term in the shift expansion in Eq.~(\ref{eq:shift_expand}) during the phase estimation measurement of $S_p$.

% kappa convention

We can also consider the effect of the self-Kerr nonlinearity in Eq.~(\ref{eq:heff2}) or the nonlinear dispersive shift $\chi' Z (a^{\dagger} a)^2$ where the strength of $\chi'/(2\pi)$ and $K/(2\pi)$ can be taken to be, say, $4$ kHz. If we represent the overall nonlinearity during the protocol as $\exp(-i \epsilon (a^{\dagger} a)^2)$ one has $|\epsilon|=K T_{\rm PE}=2 \pi \times 16 \times 10^{-3} \approx 0.1$ which is two orders of magnitude {\em larger} than $\frac{1}{n_{\rm max}^2}$ for $n_{\rm max}=25$. This means that the shift errors due to the nonlinearity will add up throughout the protocol and do not necessarily remain small correctable shift errors.

%one uses the state-independent stringent conditions in Section \ref{sec:shift} which says that $|\epsilon| n_{\max}^2 \ll 1$.
%In Section \ref{sec:shift} we also formulated a less stringent (state-dependent) condition which reads $|\epsilon| (n_{\rm max})^{3/2}\ll \sqrt{\pi}$. For $|\epsilon|=0.01$ as above and $n_{\rm max}=25$, one can barely meet this condition $|\epsilon| (n_{\rm max})^{3/2}=1.25$. This weaker condition,
%in general translates into a less stringent condition of the form $\frac{n_{\rm max}^{3/2}}{4 n_{\rm crit}} \ll 1$.

Even though the physical values of these numbers are not clad in stone, there is a simple argument which demonstrates that higher-order terms in a pertubatively-derived dispersive coupling Hamiltonian will lead to errors which require a more thorough analysis. This argument comes about from the fact that the running time of the protocol, let's say $T_{\rm round}$, is determined by $\chi \sim g^2/\Delta$, i.e. $T_{\rm round} \sim 1/\chi$. The next order in the perturbative coupling between cavity and qubit scales as $\frac{g^4}{\Delta^3}$ and we want that its overall effect during $T_{\rm round}$ to be sufficiently small, but still have a sizeable number of photons in the cavity.
We consider 
\begin{equation}
|\epsilon| n_{\rm max}^2\sim \frac{g^4}{\Delta^3} T_{\rm round} n_{\rm max}^2 \sim \frac{g^2}{\Delta^2} n_{\rm max}^2=\frac{n_{\rm max}^2}{4 n_{\rm crit}} \ll 1,
\label{eq:cond}
\end{equation}
while $n_{\rm max}$ should be $10-100$ photons in order to obtain a good code state. One has $n_{\rm max} < n_{\rm crit}$ but the stringent condition in Eq.~(\ref{eq:cond}) would require making $n_{\rm crit}$ much larger. Since $n_{\rm crit}$ should be enlarged without making the protocol last much longer (as this would enhance the strength of other sources of errors), it would be best to work at a larger detuning $\Delta$ rather than a smaller capacitive coupling between transmon qubit and cavity (coupling $g$).

One should note that this problem rapidily gets better at higher-order terms in perturbation theory. Assume that the first unwanted term comes in at $k$th order in perturbation theory, i.e. with strength $g \left(\frac{g}{\Delta}\right)^{k-1}$,  so that $T_{\rm round} g \left(\frac{g}{\Delta}\right)^{k-1} n_{\rm max}^{k/2}  \ll 1$. One then obtains the condition 
\begin{equation}
4 n_{\rm crit} \left(\frac{n_{\rm max}}{4 n_{\rm crit}}\right)^{k/2} \ll 1.
\end{equation}
For a quartic term $k=4$ this condition is quite demanding but for the next higher order term $k=6$ and $n_{\rm max} \ll 4 n_{\rm crit}$ the condition is much  more mild. \\
These arguments suggest that it is better to not treat the nonlinear dynamics as a source of errors, but rather treat it as a source of known systematic errors. One approach is to try to actively cancel their effect during the evolution so as to obtain the same code states \cite{HVS}. Another approach is to take as a given that these interactions exist and seek a code formalism that captures their effect.  Assuming that $\chi$ is tunable, one may at least hope to reduce the nonlinear dispersive shift and the self-Kerr nonlinearity in strength during the time that the dispersive interaction is off.

%One can alter the code so that its check operators $S_p$ and $S_q$ are unitarily modified in order to take into account the nonlinear dispersive shift interaction, but are still commuting along the following lines. We can take as check operators of the new code $S_p'= U^{\dagger}(\rho) S_p U(\rho)$ and $S_q'= U^{\dagger}(\rho) S_q U(\rho)$ with the nonlinear unitary interaction $U(\rho)=\exp(i \rho (a^{\dagger} a)^2)$ which are clearly still commuting.
%For this rotated code it may be possible to alter the circuit in Fig.~\ref{fig:cont_disp} so that it uses a controlled-rotation which includes the nondispersive shift.

These arguments show that further numerical analysis including the nonlinearity and open system dynamics modeled by a Lindblad equation are warranted to assess their overall effect during the phase estimation protocol.

\section{Discussion}   
In order to implement phase estimation in a shorter amount of time, it is possible to couple several transmon qubits simultaneously to a 2D (or possibly 3D) cavity. It may be interesting to consider whether a form of Shor quantum error correction (using cat state ancillas) instead of Steane quantum error correction is possible for the GKP code states. It will be worthwhile to numerically study the code state preparation protocol including the effect of nonlinearities and open system dynamics.

A somewhat different scheme for preparing code states, eigenstates of $S_p$ and $S_q$,  involves two oscillators as follows, borrowing an idea in \cite{lloyd:hybrid}. The basic idea is that by (sequentially) coupling a qubit to the oscillator to gather phase information, we are getting little information, at most one bit, per qubit used. At the same time, this transmon qubit is effectively realized as a non-linear oscillator as in Eq.~(\ref{eq:heff2}). So can one not get information faster by coupling the storage cavity to another oscillator?

Assume that we can prepare one ancilla oscillator in the state $\ket{q_a=0}$ where $q_a$ is its position quadrature (or some squeezed version thereof) and the other `storage' oscillator is in some initial state $\ket{\psi_s}=\int dp \,\alpha_p\ket{p}$. Assume an interaction between the oscillators of the form $H_{int}=p_s \otimes p_a$ for some time $t=2 \sqrt{\pi}$ so that $e^{-i H_{int} t} \ket{\psi_s}\otimes \ket{q_a=0}= \int dp \,\alpha_p \ket{p} \ket{q_a=2 p \sqrt{\pi}}$. If we can measure the $q$ quadrature of the ancilla oscillator modulo $2 \pi$, i.e. we determine $q_{\rm meas}=q_a \mod 2 \pi$, we project the system oscillator into a superposition of states with $p=\frac{q_{\rm meas}}{2 \sqrt{\pi}} \mod \sqrt{\pi}$. Shifting back this projected state depending on $q_{\rm meas}$ gives the output state  $\sum_{k=-\infty}^{\infty} \,\alpha_p \ket{p=k\sqrt{\pi}}$, thus preparing an eigenstate of $S_p$.  A similar procedure for $S_q$ on this state would then prepare the full code state. The linear coupling between the two oscillators is very natural; superconducting LC circuits with capacitive coupling realize such interaction. However, measuring a quadrature modulo $2 \pi$ is not simple: one has to make sure not to get too much information.
If the ancilla oscillator were modified to include a Josephson junction, it would make one of the quadratures, namely the phase variable, $2 \pi$-periodic and thus suitable. A rapid turning on of the Josephson junction, and thus a rapid change of the quadratic potential in $q$ or $\phi$ to a periodic cosine potential $U(\phi)=-E_J \cos(\phi)$ could freeze this state (taking $E_J   \gg E_C$, potential energy much larger than kinetic energy) and information could be read out. It is an open question whether a rapid turning-on of a Josephson junction (see e.g. \cite{BKP:protect} for a similar circuit switch idea) is experimentally feasible and what the details of such a scheme would look like.
 
\section{Acknowledgements}
We would like to thank David DiVincenzo and Robert Schoelkopf for comments and feedback. We acknowledge support through the EU via the programme SCALEQIT.

\appendix

\section{Use of Phase or Displacement Frame and Logical Gates}\label{sec:phase}

We prepare a code state or perform error correction through phase estimation of the displacement operators $S_p$ and $S_q$. The phase estimation protocol outputs an estimate for the eigenvalue $e^{i\theta_p}$ ($e^{i \theta_q}$) of $S_p$ ($S_q$), while simultaneously projecting the input state onto an approximate eigenstate $\ket{\psi_{\theta_p, \theta_q}}$ of $S_p$ and $S_q$. Error correction would then correspond to displacing this oscillator state by the corrective displacement $D_{\rm \theta_p, \theta_q}=e^{i q\theta_p/(2\sqrt{\pi})}e^{i p \theta_q/(2\sqrt{\pi})}$ such that $D_{\rm \theta_p, \theta_q}\ket{\psi_{\theta_p, \theta_q}} \approx \ket{\psi_{\theta_p=0, \theta_q=0}}$, an approximate code state. 

 Here we show that it is not necessary to do these additional displacement operations, but one can work with a phase frame similar as the Pauli frame for stabilizer codes \cite{knill:nature},\cite{BMT:review}. We will use the notation $\overline{X}=X$ and $\overline{U}$ for operators which have the action of Pauli $X$ and a unitary gate $U$ on the states in the code space. We call such operators logical operators or logical gates. In addition we write $\overline{X}_{\theta_p}$ for the Pauli $X$ operator on a code space labeled by a phase $\theta_p$ for $S_p$.\\
  
 Let us first assume that we work with perfect phase estimation and perfect code states. The operator $e^{-i\sqrt{\pi}p}$ has been called $\overline{X}$ because it maps the perfect state $\ket{0}$ onto $\ket{1}$ and vice versa. We can note that $\overline{X}$ is not hermitian, we have in fact $\overline{X}=\overline{X}^{\dagger}S_p$. The action of $\overline{X}$ and $\overline{X}^{\dagger}$ on the perfect code space is however identical, thus on the code space $\overline{X}\equiv \overline{X}^{\dagger}$ so that $\overline{X}^2=I$. If we use a code space labeled by the eigenvalue $e^{i\theta_p}$ of $S_p$, we have $\overline{X}^2=e^{i\theta_p}I$. This means that we need to redefine the logical $X$ operator on the code space characterized by eigenvalue $e^{i \theta_p}$ as $\overline{X}_{\theta_p}=D_{\theta_p,\theta_q}^{\dagger}\overline{X}D_{\theta_p,\theta_q}=\overline{X}e^{-i \theta_p/2}$ for which $\overline{X}_{\theta}^2=I$ on the $\theta_p$ code space. Similarly, $\overline{Z}_{\theta_q}=D_{\theta_p,\theta_q}^{\dagger} \overline{Z} D_{\theta_p,\theta_q}$.
Clearly, $\overline{X}_{\theta_p}$ and $\overline{Z}_{\theta_q}$ only differ from $\overline{X}$ and $\overline{Z}$ by phases so transform similarly.\\

It is known what operations are necessary to perform logical gates on the code states in the $\theta_p=0,\theta_q=0$ code space, such as the Clifford gates (the CNOT gate, the Hadamard gate, the $S$ gate) and the $T$ gate, see \cite{GKP}, as one can verify their proper action on the logical operators $\overline{X}$ and $\overline{Z}$. 

The logical Clifford group gates can be realized by linear optical transformations, i.e. linear transformations on the set of positions and momenta of $n$ oscillators, $(q_1,p_1, \ldots, q_n,p_n)$, which preserve their commutation relations. A circuit for the CNOT gate is shown in Fig.~\ref{fig:CNOT}. The Hadamard gate represents a $\pi/2$ phase delay enacting $q\rightarrow p$ and $p\rightarrow -q$. The $S={\rm diag}(1,i)$ gate enacts the transformation $q\rightarrow q$ and $p\rightarrow p-q$.

Under the action of these gates, shift errors (displacements) remain shift errors, i.e. a general displacement operator on these $n$ oscillators of the form $D=\exp(i\sum_{i=1}^n( \alpha_i p_i+\beta_i q_i))$ (with real $\alpha_i,\beta_i$) transforms to $D'= U  D U^{\dagger}$ where $D'$ has coefficients $\{\alpha_i',\beta_i'\}$. One can propagate the shift errors through a Clifford circuit:  small shift errors in several modes can add up to large shifts in one mode, similar as Pauli errors can propagate to become high-weight incorrectable Pauli errors. Furthermore, even though the Clifford gates do not amplify shift errors, shift errors inside, say, the realization of a CNOT gate can get somewhat (de-)amplified in strength as the circuit uses squeezing and squeezing acts e.g. as $p\rightarrow e^r p,q\rightarrow e^{-r} q$. \\

For logical Clifford gates $\overline{U}$ one can verify their action on the $\theta_p,\theta_q$ code space. One has 
\begin{equation}
\overline{U}\ket{\psi_{\theta_p,\theta_q}}=\overline{U} D_{\theta_p, \theta_q}^{\dagger} \overline{U}^{\dagger} \overline{U} \ket{\psi_{0,0}}=\tilde{D}_{\theta_p,\theta_q} \overline{U}\ket{\psi_{0,0}},
\end{equation}
 where $\tilde{D}_{\theta_p,\theta_p}$ is the new shift correction (displacement). This means that one never needs to do the corrective displacement, but can just keep track of the phase frame in software, 
just as for the Pauli frame in a computation with only Clifford gates.\\

When the encoding is only approximate (and we know the eigenvalues of $S_p$ and $S_q$ approximately), every code state is in principle given by a perfect code state and a distribution of shift errors, i.e. a density matrix expanded in the shift error basis as in Eq.~(\ref{eq:rho_shift}). The gates are still defined by their action on the perfect code states. The distribution of shift errors or the approximate encoding may be slightly different for every code state when these are generated, say, by the protocol in this paper. It is important to note that this does affect the proper functioning of the Clifford group gates as these gates only propagate the shift errors between logical qubits. \\

For quantum universality, one needs a $T={\rm diag}(1,\exp(i\pi/4))$ gate. This gate can in principle be realized using an ancilla $T\ket{+}$ or alternatively the ancilla $\ket{H=1}\propto S H T\ket{+}$ which is a $+1$ eigenstate of the Hadamard gate, in addition to Clifford group gates (see e.g. \cite{BMT:review}). The logical Hadamard gate $\overline{H}\colon\overline{X} \leftrightarrow \overline{Z}$ equals $\exp(i \frac{\pi}{2} a^{\dagger} a)$ which means that a $+1$ eigenstate of this operator has a photon number which is $0  \mod 4$. One can create the state $\ket{\overline{H=1}}$ by doing phase estimation for $S_p$ and $S_q$ on the vacuum state $\ket{\rm vac}$ for which $e^{i \frac{\pi}{2} a^{\dagger} a} \ket{\rm vac}=\ket{\rm vac}$ and {\rm post-selecting} on the outcomes $S_p\approx 1$ and $S_q\approx 1$. The resulting state is encoded and a $+1$ eigenstate of $\overline{H}$ as $\overline{H}$ commutes with $S_p$ and $S_q$. One can imagine that noisy ancillas $\ket{\overline{H=1}}$ are further distilled into fewer higher-quality ancillas before being used. It is important to note that a $+1$ eigenstate of $\overline{H}$ for the phase frame $\theta_q=\theta_p=0$ is not related by a displacement to a $+1$ eigenstate of $\overline{H}$ in an arbitrary phase frame $\theta_q,\theta_p$. So before one uses such ancillas one has to at least know or physically fix the phase frame that is in use. We leave a realistic method to implement $T$ (and Clifford) gates on the approximate code states in circuit-QED hardware to future work.

\section{Quantum Error Correction Using Encoded Ancillas}
\label{sec:steane_qec}

% Check whether it is possible to prepare a code state which is only an eigenstate of one of the stabilizer checks, a squeezed state for $S_q$, to use in Steane QEC

In this section we briefly review a form of quantum error correction suggested in \cite{GKP} and a simplified implementation in \cite{GK:osc}, which, for stabilizer codes, is called Steane Error Correction. In Steane Error Correction the errors are corrected by coupling the encoded data to an encoded ancilla using a CNOT gate and measuring the encoded ancilla. The advantage of using this procedure for quantum error correction is that the quantum circuit is fault-tolerant, made from linear optical elements, phase shifters and squeezers and is fully deterministic. Fault-tolerance means that shift errors are not amplified in strength by the circuit (but they do propagate) as it involves only linear optical components. A disadvantage of the method is that it requires an ancilla code state which has to be first prepared with sufficient accuracy. 

\begin{figure}[htb]
    \centering
%    \Qcircuit @C=1em @R=.7em {
%&&&&&\lstick{e^{-i u \hat{p}}\ket{\bar{\Psi}}} &\qw & \ctrl{1} & \gate{D\big(-q\mod \sqrt{\pi})
%\big)} & \qw &\ket{\bar{\Psi}}&\\
%&&&&&\lstick{\ket{\bar{+}}}& \qw & \targ & \measureD{q}\cwx{-1}
%}     
\includegraphics{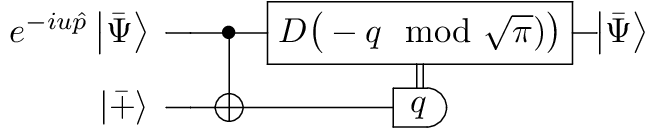}
\caption{Steane Error Correction for shifts in $q$ generated by the error shift operator $e^{-i u p}$ which are detected by the stabilizer check $S_q$. The measurement of the ancilla in the encoded $\ket{\overline{+}}$ state is a homodyne measurement of $q$. The displacement $D(-q \mod \sqrt{\pi})$ is a correction which is like a Pauli frame for a stabilizer code and does not physically need to be realized. Note that the modulo function should be taken in the interval $(-\sqrt{\pi}/2,\sqrt{\pi}/2]$. In the main text we have referred to $\ket{\overline{+}}$ as $\ket{+}$.}
\label{fig:steaneEC}
\end{figure}

The basic circuit for correcting shifts in $q$ (detected by $S_q$) is shown in Fig.~\ref{fig:steaneEC}. A similar circuit holds for $S_p$ where the encoded ancilla is prepared in $\ket{\overline{0}}$ (denoted as $\ket{0}$ in the main text). The CNOT gate (a linear optical gate) has the following effect on the two quadratures of control (c) and target (t) mode:  $q_c \rightarrow q_c, q_t \rightarrow q_c+q_t,p_c \rightarrow p_c-p_t,p_t \rightarrow p_t$.
The CNOT gate cannot be implemented with only beamsplitters and phase-shifters as the symplectic transformation matrix of the CNOT, i.e. the matrix applied to the vector $(q_c,p_c,q_t,p_t)$ is not orthogonal, while it is always orthogonal for any passive linear optics transformation, see e.g. \cite{weedbrook:RMP}.  The CNOT gate can be realized by beamsplitters and squeezing \cite{braunstein:squeeze}. The optimal CNOT circuit expressed in terms of these elementary components is depicted in Fig.~\ref{fig:CNOT}. \\

In the Glancy and Knill method in \cite{GK:osc} the CNOT gate which uses two beam-splitters and two squeezers is replaced by a single beam-splitter and a single (but stronger) squeezer. The $\sqrt{\pi}/6$ shift error bound is arrived at by arguing how shift input errors on data and ancilla propagate through the quantum error correction circuit leading to shift errors which have to be corrected in the next round.

Basically a shift error $\exp(-i u p)\exp(-i v q)$ that needs to be properly corrected in steady state (which requires $|u|,|v| < \sqrt{\pi}/2$) results from the propagation of three separate shift errors $u=u_I+u_p+u_q,v=v_I+v_p+v_q$, which are the initial errors $u_I,v_I$ and the errors that occur somewhere in the quantum error correction circuits for $S_p$ ($u_p,v_p$) and $S_q$ ($u_q,v_q$), hence all $|u_i|,|v_i| < \sqrt{\pi}/6$.
Thus if all initial shift errors on encoded data and ancilla state are of strength at most $|u|,|v| <\sqrt{\pi}/6$, errors remain correctable through the repeated application of these circuits, assuming that the linear optical circuits themselves are faultless. If the linear optical circuits are faulty, one has to propagate the shift errors that occur during the circuit forward which will result in a somewhat lower threshold value. 

% BMT this argument holds both for basic Steane QEC here as well as the GK modification

\begin{figure}[htb]
    \centering
    \includegraphics[width=\hsize]{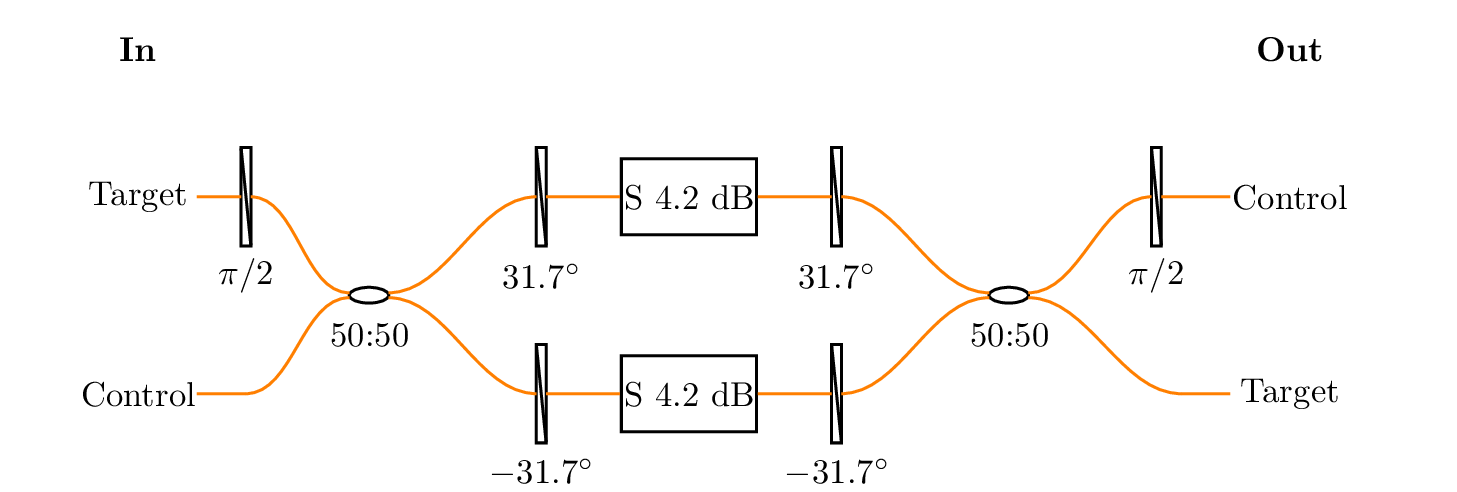} \caption{(Color online) A CNOT gate on two bosonic modes, target and control. $S$ denotes squeezing and its strength in dB (see squeezing convention below Eq.~(\ref{eq:approx})). The phase shifters shown explicitly can of course be absorbed in the action of the squeezers or the 50:50 beam-splitters.}
\label{fig:CNOT}
\end{figure}

\begin{figure}[htb]
    \centering
    \includegraphics[width=\hsize]{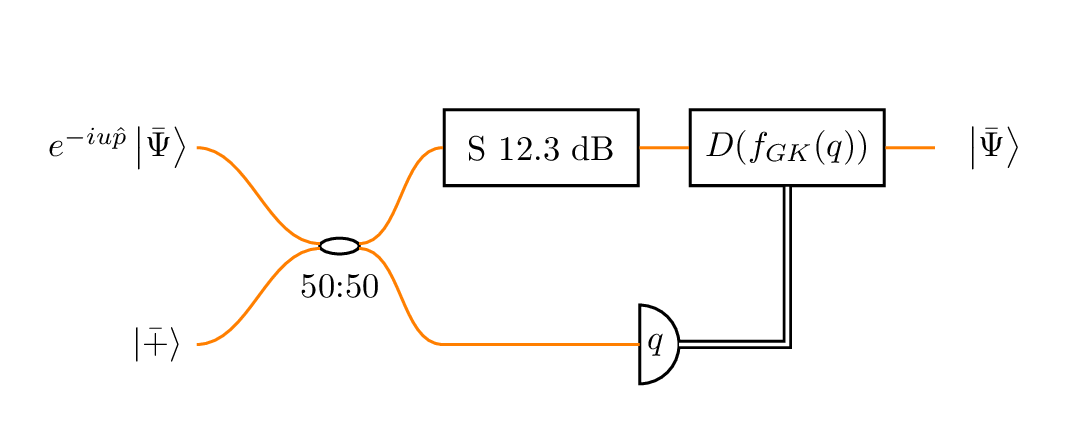} \caption{%DW RESUBMIT CHANGE: added color online
    (Color online) Steane Quantum Error Correction using the method of Glancy and Knill for shifts $e^{-i u p}$ which are detetected by the stabilizer check $S_q$. The circuit only uses a single beam-splitter, a homodyne measurement of $q$ and much stronger squeezing followed by a different corrective displacement $D(f_{GK}(q))$ which is explicitly given in \cite{GK:osc}.}
\label{fig:GKsteaneEC}
\end{figure}

 \section{Choice of Feedback Phases}
\label{sec:adap_phase}

Let us assume that the feedback phases $\varphi_1,\ldots,\varphi_{m-1}$ have been fixed and the bit string $x[m-1] \equiv\{ x_1\ldots x_{m-1}\}$ has been generated. How do we choose the next phase $\varphi_m$ in an optimal way? Using conditional probability distributions (i.e. the identities $P(A|BC)=P(C|AB) P(A|B)/P(C|B)$ and $P(C|AB)=P(CB|A)/P(B|A)$), the probability distribution for $\theta$ after obtaining the measurement result  $x[m]=\{x_m, x[m-1]\}$ equals:
\begin{align}
P_{\varphi}(\theta|x[m]) &= \frac{ P(\theta|x[m-1]) P_{\varphi_m}(x_m|\theta,x[m-1])}{P_{\varphi}(x_m|x[m-1])} \notag \\
&= \frac{ P(\theta|x[m-1]) P_{\varphi}(x[m]|\theta)}{P_{\varphi}(x_m|x[m-1])P(x[m-1]|\theta)}.
\label{eq:condition}
\end{align}
Here we have explicitly kept the dependence of probability distributions on the phase $\varphi$ which is chosen for the $m$th qubit measurement as this is the quantity that one wants to optimize over. 
This phase $\varphi$ is chosen as the one which maximizes the sharpness of the a posterior probability distribution $P_{\varphi}(\theta|x[m])$ averaged over possible outcomes $x_m$ (given the previous outcomes and choices for phases), or using Eq.~(\ref{eq:sharpness}),
\begin{align}
\varphi_m &= \argmax_{\varphi} \sum_{x_m=0,1} P_{\varphi}(x_m| x[m-1]) S[P_{\varphi}(\theta|x[m])] \notag \\
&=\argmax_{\varphi}\sum_{x_m=0,1}\left|\int d \theta e^{i\theta} \frac{P(\theta|x[m-1]) P_{\varphi}(x[m]|\theta)}{P(x[m-1]|\theta)} \right|.
\label{eq:opt}
\end{align}

The expression for $P_{\varphi}(x[m]|\theta)$ is simple as the measurement results of each round are independent, i.e. 
\begin{equation}
P_{\varphi=\varphi_m}(x[m]|\theta)=\prod_{i=1}^m P_{\varphi_i}(x_i|\theta)=\prod_{i=1}^m \cos^2\left(\frac{\theta+\varphi_i}{2}+x_i \frac{\pi}{2}\right).
\label{eq:prod2}
\end{equation}

If we assume no prior knowledge on $\theta$ one can show that $P(\theta|x[m-1]) \propto P(x[m-1]|\theta)$ where the proportionality constant $c$ is independent of $\theta$ and $\varphi_m$ (but may depend on previous phases $\varphi_{m-1}$ etc.). The argument is as follows. For the first measurement we write $P_{\varphi_1}(\theta|x_1)=P_{\varphi_1}(x_1|\theta) P(\theta)/P(x_1) \propto P_{\varphi_1}(x_1|\theta)$ if we assume that $P(\theta)$ is a flat distribution (we have no prior knowledge about $\theta$) and $P(x_1)=1/2$. Using Eq.~(\ref{eq:condition}) we see that this then holds for for all measurement outcomes by induction, i.e. $P(\theta|x[m-1]) \propto P_{\varphi_m}(x[m-1]|\theta)/P_{\varphi_{m-1}}(x_{m-1}|x[m-2])$.

This implies that the optimization in Eq.~(\ref{eq:opt}) is equivalent to 
\begin{equation}
\varphi_m= \argmax_{\varphi} \sum_{x_m=0,1}\left|\int d \theta e^{i\theta}P_{\varphi}(x[m]|\theta) \right|  \nonumber,
\end{equation}
which can also be rewritten as 
\begin{widetext}
\begin{equation}
\varphi_m=\argmax_{\varphi} \sum_{x_m=0,1} \left|\int d\theta e^{i \theta} \cos^2\left(\frac{\theta+\varphi}{2}+x_m\frac{\pi}{2} \right)\prod_{j=1}^{m-1} \cos^2\left( \frac{\theta+\varphi_i}{2}+x_i\frac{\pi}{2} \right) \right|.
\label{eq:feedback2}
\end{equation}
\end{widetext}

\section{Direct Controlled-Displacement}
\label{sec:direct}

% This means that $\omega_q(t)=\omega_q$ during this time-interval $T$, but can be different outside this interval.

We model the effect of applying a microwave drive to the cavity mode without including cavity decay. The reason for not including cavity decay is that the parameters of the protocol should be chosen to work without cavity decay such that the loss of photons from the cavity acts as a source of (shift) errors during this ideal protocol.

The drive field is represented as a term $H_{\rm drive}(t)=\lambda(t) (a + a^{\dagger})$ where $\lambda(t)=\Omega_x(t) \cos(\omega_d t)+\Omega_y(t) \sin(\omega_d t)$ with $\omega_d$ the drive frequency and $\Omega_x(t), \Omega_y(t)$ the envelope of the pulse. The drive frequency $\omega_d$ is chosen as $\omega_d=\omega_r+ \chi$. The total Hamiltonian $H_{\rm tot}(t)=H_{\rm eff}+ H_{\rm drive}(t)$ (in the lab frame) with $H_{\rm eff}$ in Eq.~(\ref{eq:heff}) equals
\begin{equation}
H_{\rm tot}(t)= (\omega_r I-\chi(t) Z)a^{\dagger} a+\lambda(t) (a + a^{\dagger})- \frac{\omega_q Z}{2}.
\label{eq:hcouple}
\end{equation}

In order to analyze the effect of the dynamics due to $H_{\rm tot}(t)$, we use the following useful tool/fact. For a Hamiltonian of the form $H(t)= \omega a^\dag a + \lambda(t) (a+a^\dag)$, the unitary time evolution can be written as a product of an overall rotation, an overall displacement and a phaseshift, i.e. 
\begin{equation}
\mathcal{T} e^{-i \int_{0}^{T} dt' H(t')} = R(\omega T) D(\gamma) \exp(i \Psi),
\end{equation}
with $\gamma=-i \int_{0}^{T} dt' \lambda(t') e^{-i \omega t'}$ and  $\Psi= \int_0^T dt \int_t^{T} dt' \lambda(t) \lambda(t') \sin(\omega (t'-t))$. We have $R(\omega T)=\exp(-i \omega T a^{\dagger} a)$. This equality can be derived using the Suzuki-Trotter decomposition:

\begin{align}
&\mathcal{T} e^{-i \int_{0}^{T} dt' H(t')}= \lim_{n \rightarrow \infty} \prod_{j=1}^n\left(\underbrace{e^{\frac{-iT}{n}\lambda(t_n) (a+a^\dag)}}_{D_{t_j}}\underbrace{e^{\frac{-i\omega T}{n} a^\dag a}}_{R_n}\right)\nonumber \\
&= \lim_{n \rightarrow \infty} R_n^n (R_n^{-n}D_{t_n} R_n^n) \ldots (R_n^{-2} D_{t_2} R_n^2) (R_n^{-1} D_{t_1} R_n)\nonumber \\
&=R(\omega T)  D\left(\lim_{n \rightarrow \infty}\sum_{j=1}^n \frac{-i \lambda(t_j) T}{n} e^{-i\omega T \frac{j}{n}}\right) e^{i\Psi},  \nonumber
\end{align}
where $\Psi$ is determined using $D(\alpha) D(\beta)=\exp(i {\rm Im}(\alpha\beta^*))D(\alpha+\beta)$ and taking the limit $n \rightarrow \infty$. 

We can apply this equality to the cavity-transmon dynamics $V(0,T) \equiv \mathcal{T}\exp(-i \int_0^T dt' H_{\rm tot}(t'))$ as it is diagonal in the $\{\ket{0},\ket{1}\}$ basis: 
\begin{align}
V(0,T)&=R((\omega_r-\chi)T)D(\gamma_-) e^{i({\Psi_-} +\omega_q T/2)}\ket{0}\bra{0}\notag \\
&+R((\omega_r+\chi)T)D(\gamma_+) e^{i({\Psi_+}-\omega_q T/2)}\ket{1}\bra{1},  \nonumber
\end{align}
where $\Psi_{\pm}= \int_0^T dt \int_t^{T} dt' \lambda(t) \lambda(t') \sin((\omega_r\pm \chi) (t'-t))$ and $\gamma_{\pm}=-i \int_{0}^{T} dt' \lambda(t') e^{-i (\omega_r \pm \chi)t'}$, the Fourier transform of $\lambda(t)$ at frequencies $\omega_r\pm \chi$. We note that the relative rotation when qubit is in state $\ket{1}$ versus $\ket{0}$ is $R(2\chi T)$. Thus without constraining $\chi T$, the interaction between qubit and cavity field realizes a controlled-rotation in addition to the desired controlled-displacement transformation $D(\gamma_{\pm})$.  We can understand the undesired effect of the relative rotation in the sketch in Fig.~\ref{fig:phasespace_sketch} in which a controlled-displacement gate (and a some relative controlled-rotation) is applied on, say, a cat state.

If we demand that $T$ is chosen such that $\chi T=\pi k$, the relative rotation acts trivially $R(2 \chi T)=I$. One thus has to choose $T=\pi/\chi$ which for $\chi/2\pi=2.5$ MHz (which is the same as $\chi/2\pi$ in the experiment in \cite{vlastakis+:cat100}), equals $200$ nsec.

\begin{figure}[htb]
   \centering
\subfigure[][]{
\includegraphics[width = .45\hsize]{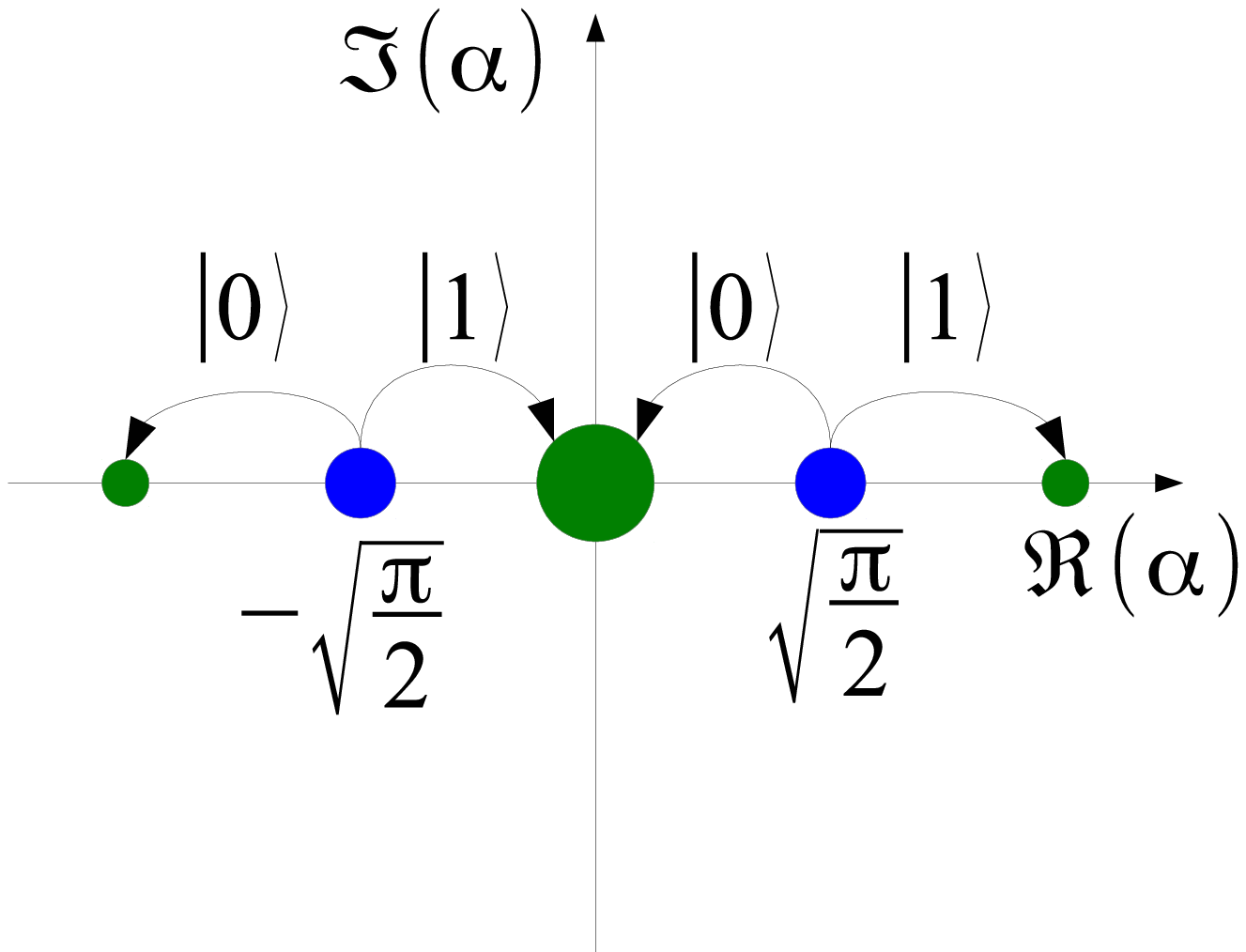}
}
\subfigure[][]{
\includegraphics[width = .45\hsize]{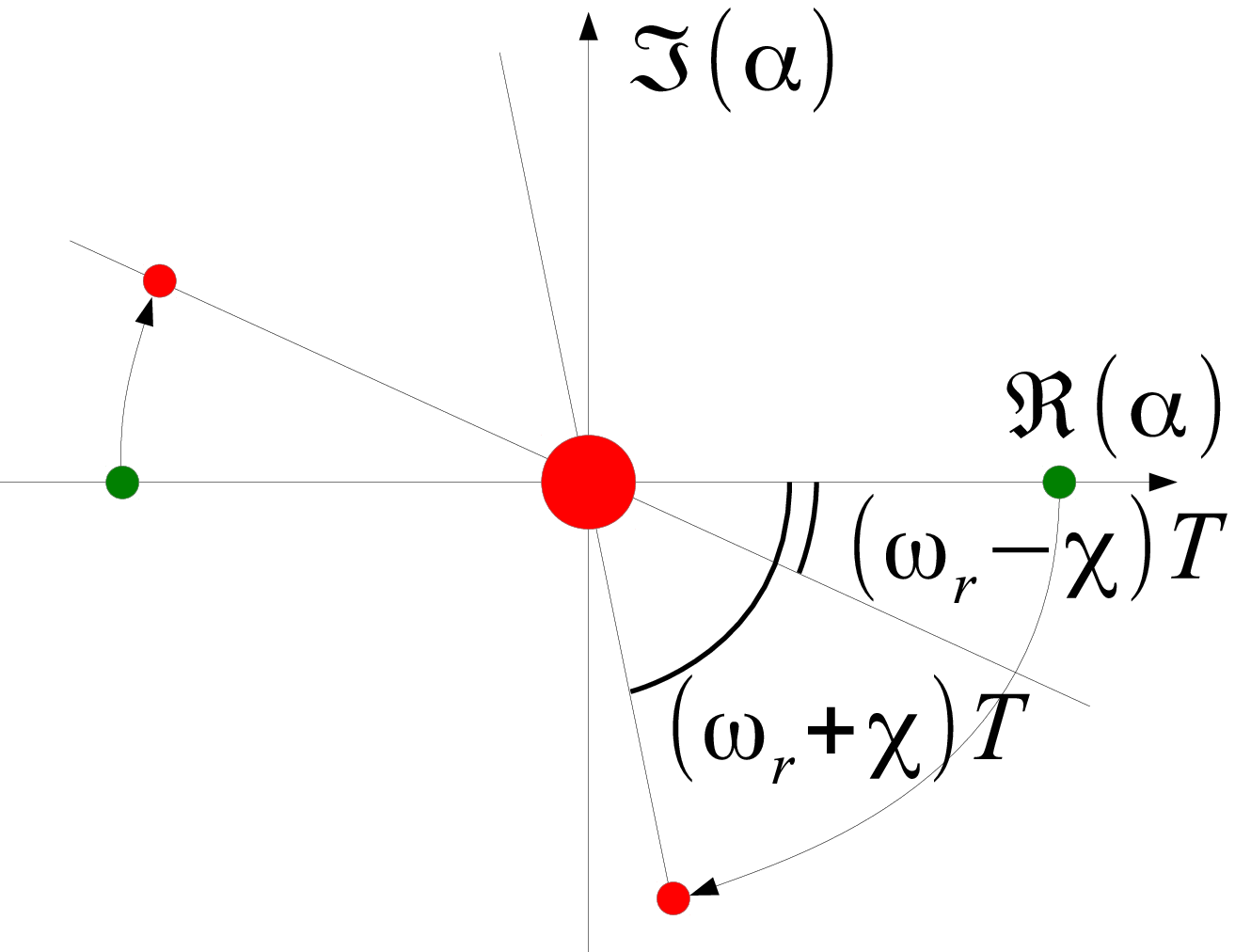}
}
   \caption{(Color online) Phase space sketches: the size of the blobs represents amplitude not quadrature uncertainty. (a) A cat state $\ket{\alpha}+\ket{-\alpha}$ with $\alpha=\sqrt{\pi/2}$ (blue dots) gets displaced by $\pm \sqrt{\pi/2}$ depending on a qubit state $\ket{0}$ or $\ket{1}$ along the $\Re(\alpha)$ line producing the green dots. Assuming constructive interference of phases, the amplitude for $\ket{\rm vac}$ is double that of the outlying states. Repetition of this scheme with constructive interference produces a binomial distribution of coherent states according to a Pascal triangle. In the phase estimation protocol, the input state is not a cat state but the squeezed vacuum (along the $\Re(\alpha)$-direction) and the protocol generates superpositions of displaced squeezed vacua on a line. In (b) it is shown how an additional phase space rotation which is different for $\ket{0}$ as for $\ket{1}$ produces a superposition of three states (red dots) which are no longer on a line, hence will not represent a code state. When the condition $2\chi T=2\pi$ is met, this relative unwanted rotation vanishes. 
   }
    \label{fig:phasespace_sketch}
\end{figure}

We can write down the expressions for the displacements $D(\gamma_{\pm})$. For a square displacement pulse (with fictitious zero rise time) which is turned on for the entire interval $T$, the displacements $D(\gamma_\pm)$ are given by
\begin{align}
\gamma_+&=\frac{(-i \Omega_x- \Omega_y)T}{2}+\delta_+,\notag \\
\gamma_-&=\frac{(-i \Omega_x-\Omega_y)T}{2}e^{-i \omega_r T} {\rm sinc}(\omega_r T)\notag \\
&+\frac{(-i \Omega_x+\Omega_y)T}{2} e^{i \chi T}{\rm sinc}(\chi T),
\end{align}
with ${\rm sinc}(x)=\sin(x)/x$ and a displacement error $\delta_+=e^{-i (\omega_r+\chi) T}\frac{T(-i \Omega_x-\Omega_y)}{2}{\rm sinc}((\omega_r+\chi) T)$.  We wish that $\gamma_- \approx 0$ (cavity is off-resonance) and note that this depends, in principle, on ${\rm sinc}(\chi T)$ and ${\rm sinc}(\omega_r T)$. One can alway bound $|{\rm sinc}(x)| \leq 1/|x|$, which shows that the dominant term in the expression for the off-resonant displacement $\gamma_-$ scales in strength like $\frac{|\Omega_x+i \Omega_y|T}{ 2\chi T}$. However, for our chosen pulse length $T=\pi/\chi$, ${\rm sinc}(\chi T)=0$ and this analysis is too pessimistic.  

For $\omega_r/2\pi=10$ GHz (for example) one has $\frac{\omega_r T}{2 \pi}=\frac{\omega_r}{2\chi}=2 \times 10^{3}$ so that $|{\rm sinc}(\omega_r T) |   < 0.5 \times 10^{-3}$. For the well-justified approximation $\omega_r \gg \chi$,  $|\delta_+| \approx |\gamma_-| \approx  |\gamma_+| {\rm sinc}(\omega_r T) < |\gamma_+| \times 0.5 \times  10^{-3}$. This means that the off-resonant displacement is a factor $10^3$ less than the on-resonant displacement. 

For the $S_p$ measurement, one can then choose $\Omega_x=0$ and $\gamma_+=\sqrt{2\pi}=\frac{-\Omega_y T}{2}$ (assuming $\gamma_-\approx 0$).
 Similarly, for the $S_q$ measurement one can choose $\Omega_y=0$ etc. The value of the parameter $\Omega_y$ (or $\Omega_y$) depends on the external coupling of the cavity with the drive-line as well as the applied microwave power, e.g. $\Omega_y \propto \sqrt{\kappa_{\rm ext} F_t}$ where $F_t$ is the photon flux per unit time. For the square pulse this photon flux per unit time is assumed to be constant during the interval $T$. Thus $P \propto \omega_d  T F_t= \omega_d n_{\rm pulse} $ is the total input power and $n_{\rm pulse}$ the total number of photons in the displacement pulse. 
We thus find that one can enact the controlled-displacement gate when $2 \pi \approx \kappa_{\rm ext} n_{\rm pulse} T$.  Given that $T$ is fixed, depending on $\chi$, and $\kappa_{\rm ext}$ is required to be small in order to have a high-Q cavity, this shows that by increasing $n_{\rm pulse}$ or the total input power one can always achieve the desired displacement.

It is important to note that if the condition $T=\pi/\chi$ is not accurately fulfilled, and one uses the pessimistic bound $|{\rm sinc}(x)|\leq \frac{1}{x}$, the relative difference in displacement equals $|\gamma_-| /|\gamma_+| \sim \frac{1}{\chi T}=\frac{1}{\pi}$. Also, it is not required that $\gamma_- \approx 0$ as long as the value of the {\em relative} displacement (for, say, $S_p$) is large enough, i.e. $\gamma_+-\gamma_-=\sqrt{2\pi}$ and the values for $\gamma_{\pm}$ are known (calibrated). This means that we do not need to fullfill the condition $T_{\rm pulse}=T=\pi/\chi$ exactly. What happens when $T_{\rm pulse}  < T$ and we also include the finite rise-time of the pulse? The expressions derived above will remain valid, but lead to a different $\gamma_{\pm}$ (and $\Psi_{\pm}$) which could be fine-tuned using the pulse-shape $\Omega_x(t), \Omega_y(t)$. These values of $\gamma_{\pm}$ can also be calibrated by testing the controlled-displacement gate on the vacuum state. \\

Let us comment on the phases $\Psi_\pm$ which affect the transmon qubit and thus need to be taken into account if the qubit undergoes further rotations. Obviously, only the relative phase $\Psi_+-\Psi_-$ will affect the qubit state. If this relative phase remains {\em fixed} for every round of the $S_p$ phase estimation measurement (and possibly different but fixed for the $S_q$ measurement), it will not affect the accuracy of the prepared code states, even if this relative phase is unknown. It essentially means that whenever one does controlled-$S_p$ in the protocol, one implements controlled-$S_p e^{i(\Psi_+-\Psi_-)}$ instead. If the phase difference is not fixed, but known (so it represents a systematic error), it can be corrected with a rotation of the qubit. A fluctuating phase difference $\Psi_+-\Psi_-$ will however lead to a noisier estimate in the phase estimation algorithm, hence shift errors in the resulting code state.

\begin{figure*}[htb]
\centering
\includegraphics[width = \textwidth]{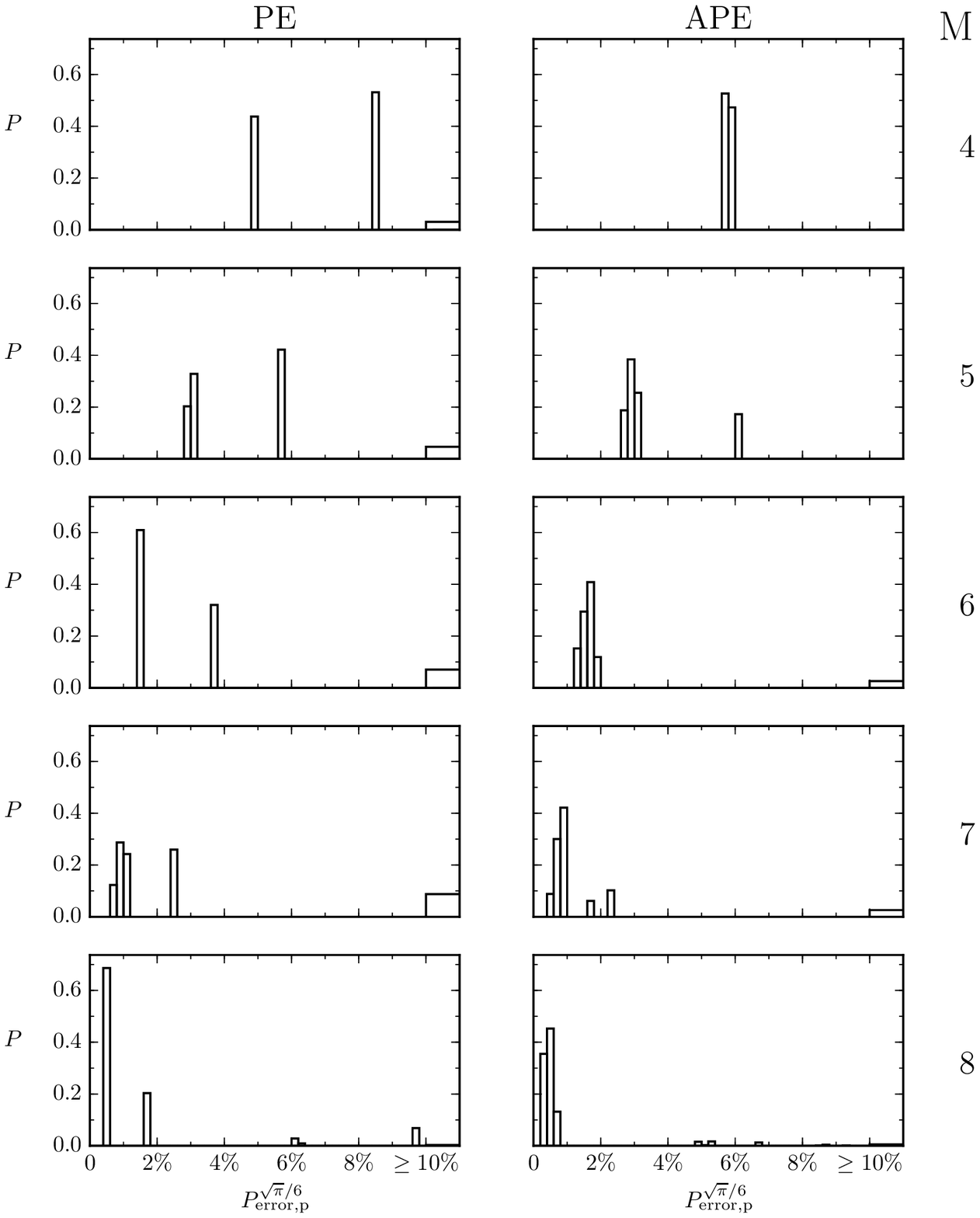}
\caption{
%DW RESUBMIT CHANGE: Figure now includes proper weighing of states. As a result, I have restricted the figure to the range 0-10%. 
%QUESTION: Are the tick marks ok? P is currently 0-1, while P_error is 0-100\%. This is inconsistent, but I think less cluttered. The code is already written, so I could do it differently in about 10 minutes. Changing the location of tickmarks (except the large bin) can also be done easily.
Probability $P$ to prepare a code state with an effective error rate $P_{\text{error,p}}^{\sqrt{\pi}/6}$ using $M$ ancilla qubits (rounds) and an infinitely squeezed vacuum state as initial state, binned to $0.2\%$. Errors due to finite squeezing are uncorrelated to the effective error rate shown here and can be taken into account using Fig. \ref{fig:squeeze}. The wide bin $P_{\text{error,p}}^{\sqrt{\pi}/6}\geq 10\%$ contains all measurement outcomes with effective error rate $\geq 10\%$. Left: non-adaptive phase estimation of $S_p$. Right: adaptive phase estimation of $S_p$.}
\label{fig:ape_hist}
\end{figure*}

\bibliography{cat_refs}

\end{document}